\documentclass[fleqn,usenatbib,onecolumn]{mnras}
\usepackage{newtxtext,newtxmath}

\usepackage[T1]{fontenc}
\usepackage{ae,aecompl}

\usepackage{epsfig}
\usepackage{graphicx}
\usepackage{amsmath}
\usepackage{amssymb}
\usepackage{mathrsfs}
\usepackage{mathtools}
\usepackage{hyperref}
\usepackage[dvipsnames]{xcolor}

\title[Instability of non-Keplerian warped discs]{Instability of non-Keplerian warped discs}
\author[Do\u{g}an \& Nixon]{
S.~Do\u{g}an$^{1,2}$\thanks{suzan.dogan@ege.edu.tr} \& C.~J.~Nixon$^{2}$\thanks{cjn@leicester.ac.uk }\vspace{0.1in}\\
$^{1}$Department of Astronomy and Space Sciences, University of Ege, Bornova, 35100, ${\dot {\rm I}}$zmir, Turkey\\
$^{2}$Department of Physics and Astronomy, University of Leicester, Leicester, LE1 7RH, UK\\
}

\date{Draft version, \today.}
\pubyear{2020}
\pagerange{\pageref{firstpage}--\pageref{lastpage}}
\begin{document}
\label{firstpage}
\maketitle

\begin{abstract}
  Many accretion discs are thought to be warped. Recent hydrodynamical simulations show that (i) discs can break into distinct planes when the amplitude of an imposed warp is sufficiently high and the viscosity sufficiently low, and that (ii) discs can tear up into discrete rings when an initially planar disc is subject to a forced precession. Previously, we investigated the local stability of isolated, Keplerian, warped discs in order to understand the physics causing an accretion disc to break into distinct planes, finding that anti-diffusion of the warp amplitude is the underlying cause. Here, we explore the behaviour of this instability in disc regions where the rotation profile deviates from Keplerian. We find that at small warp amplitudes non-Keplerian rotation can stabilize the disc by increasing the critical warp amplitude for instability, while at large warp amplitudes non-Keplerian rotation can lead to an increased growth rate for discs that are unstable. Tidal effects on discs in binary systems are typically weak enough such that the disc remains close to Keplerian rotation. However, the inner regions of discs around black holes are strongly affected, with the smallest radius at which the disc can break into discrete planes being a function of the black hole spin. We suggest that interpreting observed frequencies in the power spectra of light curves from accreting compact objects as nodal and apsidal precession of discrete orbits requires an instability that can break the disc into discrete rings such as the one explored here. 
\end{abstract}


\section{Introduction}
Accretion discs are generally warped, either during formation from turbulent initial conditions in star formation \citep[e.g.][]{Bate:2010aa} and supermassive black hole feeding \citep[e.g.][]{Lucas:2013aa}, or through processes which warp an initially planar disc such as Lense-Thirring precession from a spinning black hole \citep{Lense:1918aa, Bardeen:1975aa}, the gravitational forcing from a binary companion \citep{Papaloizou:1995ab}, magnetically driven warping by an inclined dipole \citep{Lai:1999aa} or warping with a planet on an inclined orbit \citep[e.g.][]{Xiang-Gruess:2013aa}. Warped disc profiles have been directly observed in various systems, e.g. in the maser disc found in the galaxy NGC 4258 \citep{Miyoshi:1995aa}, in the debris disc around $\beta$ Pic \citep{Heap:2000aa}, and indirectly inferred from long-term periodicities in low-mass X-ray binaries, e.g. Her X-1 \citep[e.g.][]{Katz:1973aa} and precessing jets, e.g. in the high-mass X-ray binary SS 433 \citep{Margon:1984aa}. Warps are now routinely observed in protoplanetary discs with e.g. ALMA \citep[e.g.][]{Casassus:2015aa}.

The local thermal-viscous stability of a warped and isolated (i.e. no external torque) disc was first explored by \cite{Ogilvie:2000aa}, who concluded that the disc becomes unstable when the warp amplitude is sufficiently large and the viscosity is sufficiently small. In \cite{Dogan:2018aa} this was investigated in more detail with the aim of identifying why discs can break into distinct planes \citep[as found in the simulations of, e.g.,][]{Larwood:1996aa,Fragner:2010aa,Lodato:2010aa,Nixon:2012aa,Nixon:2012ad,Nixon:2013ab,Facchini:2013aa,Dogan:2015aa,Nealon:2015aa,Liska:2020aa}. The disc becomes unstable when the diffusion rate of the warp is not maximal at the location of the maximum warp amplitude, and therefore anti-diffusion of the warp drives a sharp break between two parts of the disc. The warp amplitude in this region grows and the surface density drops, leading to rings of matter at different radii which occupy different orbital planes (which, in the case that the disc warp amplitude grows with time due to a forced differential precession of the disc orbits, leads to disc tearing; \citealt{Nixon:2012ad}\footnote{We note that recent observational evidence for this process occurring in a protoplanetary disc is presented by \cite{Kraus:2020aa}.}). In the case of a planar disc this unstable behaviour reduces to the Lightman-Eardley viscous instability which leads to discontinuities in the surface density of the disc \citep[i.e. rings of matter which occupy the same orbital plane;][]{Lightman:1974aa}.

In a warped disc, the local orbital plane of the matter changes with the disc radius. The misalignment between the high-pressure midplane regions of two neighbouring rings produces a pressure gradient in the radial direction. Therefore, a fluid element orbiting in the disc with a local orbital frequency $\Omega$ encounters a radial pressure force which oscillates around the orbit at the same frequency $\Omega$. This induces epicyclic motion within the disc. When the disc is Keplerian, the orbital and the epicyclic frequencies are the same, i.e. $\Omega = \kappa$ where $\kappa$ is the epicyclic frequency, creating a resonance between the forcing and the epicyclic motion, and this excites strong epicyclic motions which communicate the disc warp in the radial direction. However, the disc orbital and epicyclic frequencies need not be the same.  In many astrophysical systems, the disc is subjected to additional effects (arising from, e.g., the gravitational field of a spinning black hole or the gravitational potential of a binary) leading to departures from Keplerian rotation. Non-Keplerian rotation weakens the resonance between the forcing and the motion. In this case, the numerical values of the diffusion coefficients in the warped disc equations \citep[see][]{Ogilvie:1999aa,Ogilvie:2000aa,Ogilvie:2013aa}, and thus the disc evolution, will not be the same as for a Keplerian disc.

In this work, we explore the influence of non-Keplerian rotation on the instability of warped discs described by diffusive propagation of the warp. In Section 2 we discuss the systems where departures from Keplerian rotation are expected to be important. In Section 3 we provide, for a range of rotation profiles, numerical evaluations of the diffusion coefficients, the growth rates of the instability and the critical warp amplitudes at which the disc becomes unstable. In Section 4 we present our conclusions.

\section{Non-Keplerian Rotation}
In the Newtonian regime, the radial epicyclic frequency for an accretion disc is defined by 
\begin{equation}
\label{eq:kappa}
{\kappa}^2=4\Omega^2 + \frac{{\rm d}\Omega^2}{{\rm d} \ln r}
\end{equation}
where $\Omega(r)$ is the angular frequency, and $\Omega(r)=\big(GM/r^3\big)^{1/2}$ for a Keplerian disc. As in \cite{Ogilvie:2000aa}, it is useful to introduce the dimensionless rate of orbital shear $q$ for the stability analysis which will be presented in Section \ref{sec:analysis}:
\begin{equation}
\label{eq:q}
q=-\frac{\rm d \ln \Omega}{\rm d \ln r}.
\end{equation}
Therefore, the Newtonian dimensionless shear rate and the epicyclic frequency can be related to each other by
\begin{equation}
\label{eq:kappa_q}
\tilde{\kappa}^2=4-2q
\end{equation}
where $\tilde{\kappa} = \kappa/\Omega$ is the dimensionless epicyclic frequency. For Keplerian orbits $q = 3/2$, and $\tilde{\kappa}^2 = 1$. In many astrophysical systems the rotation is not strictly Keplerian with the dimensionless rate of shear higher than 3/2, and thus $\tilde{\kappa}^2 < 1$. In this section we present some cases where non-Keplerian rotation is expected to be important: (1) discs subject to tides from a binary, i.e. circumprimary/circumsecondary and circumbinary discs, and (2) discs around a (spinning or Kerr) black hole. We aim to determine the range of possible deviations in the rate of orbital shear and the epicyclic frequency from the Keplerian case.

\subsection{Discs with a tidal torque from a binary}
\subsubsection{Circumstellar discs}
\label{sec:nonkep2}
In a binary system, a disc around one component can be perturbed by the gravitational (tidal) field of the companion, leading to non-closed disc orbits (described by apsidal and nodal precession) with $\Omega \ne \kappa$. The perturbation is strongest in the outer regions of the disc closest to the companion. The strength of the deviation depends on the companion's mass and the disc radius.

\begin{figure}
  \begin{center}
        {\includegraphics[angle=270,scale=.43]{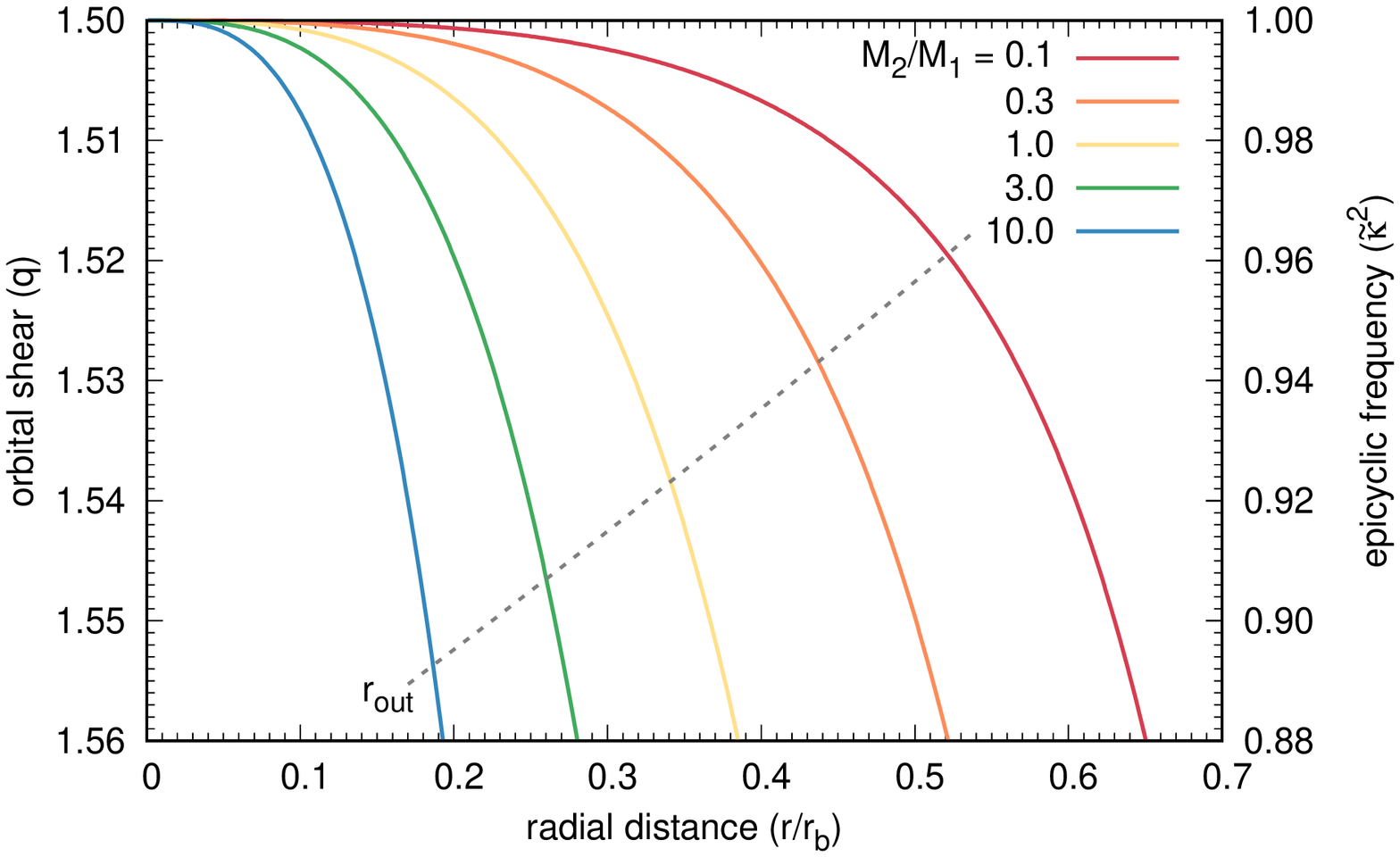}}
        {\includegraphics[angle=270,scale=.43]{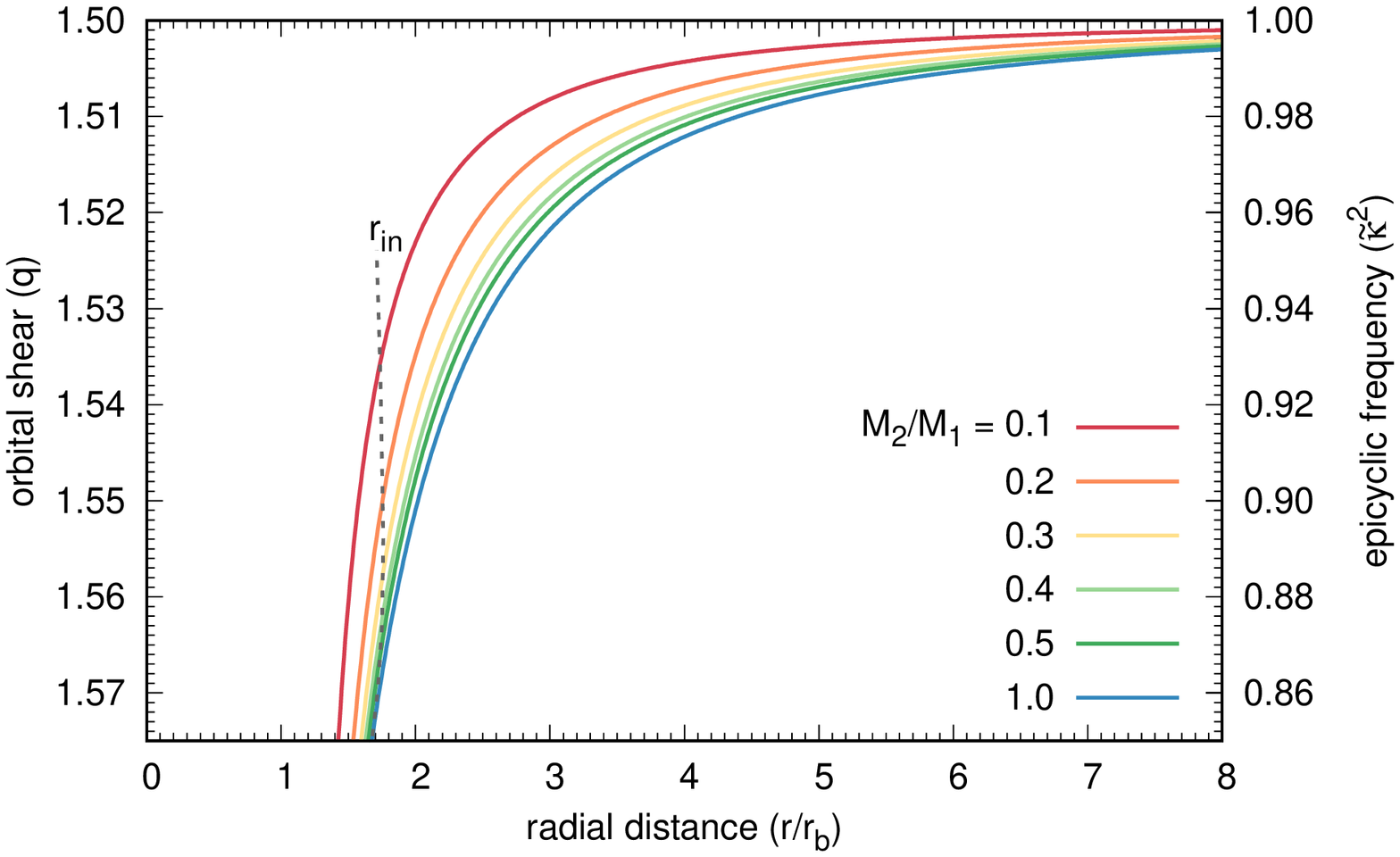}}
              \caption{Left-hand panel: This shows the radial distribution of the dimensionless orbital shear rates and corresponding dimensionless epicyclic frequencies for circumstellar discs for various mass ratios. The dashed line approximates the tidal radius which determines the outer edge location of the disc and the corresponding $q$ value. The disc cannot extend beyond this distance. Right-hand panel: This shows the radial distribution of the dimensionless orbital shear rates and corresponding dimensionless epicyclic frequencies for circumbinary discs for various mass ratios. The inner edge of the disc predicted by \protect\cite{Artymowicz:1994aa} and corresponding $q$ values are shown with the dashed line. We note that in the case of a circumbinary disc the values of $q$ and $\tilde{\kappa}^2$ are identical for mass ratio $M_2/M_1$ and $M_1/M_2$.}
        \label{fig:cp-cb}
  \end{center}
\end{figure}

The orbital and the epicyclic frequencies for a circumstellar disc are given by \citep[e.g.][]{Lubow:2000aa}
\begin{equation}
\label{eq:cs_omega}
\Omega^2=\frac{GM_1}{r^3}+\frac{GM_2}{2r_{\rm b}^2r}\bigg[\frac{r}{r_{\rm b}}b^{(0)}_{3/2}\bigg(\frac{r}{r_{\rm b}}\bigg)-b^{(1)}_{3/2}\bigg(\frac{r}{r_{\rm b}}\bigg)\bigg]
\end{equation}
and
\begin{equation}
\label{eq:cs_kappa}
\kappa^2=\frac{GM_1}{r^3}+\frac{GM_2}{2r_{\rm b}^2r}\bigg[\frac{r}{r_{\rm b}}b^{(0)}_{3/2}\bigg(\frac{r}{r_{\rm b}}\bigg)-2b^{(1)}_{3/2}\bigg(\frac{r}{r_{\rm b}}\bigg)\bigg]
\end{equation}
where
\begin{equation}
b_{\gamma}^{(m)}(x)=\frac{2}{\pi}\int_0 ^\pi \cos(m\phi)(1+x^2-2x \cos \phi )^{-\gamma} d\phi
\end{equation}
is the Laplace coefficient, $r_{\rm b}$ is the binary separation and $M_1$ is the mass of the component which has a disc (for simplicity we have assumed a circular binary). In Fig. \ref{fig:cp-cb} (left-hand panel) we show $q$ and $\tilde{\kappa}^2$ as a function of radius ($r/r_{\rm b}$) for various values of $M_2/M_1$. As the deviation from Keplerian rotation increases with increasing radius in the disc, we need to estimate the outer radius of the disc to find the maximal deviation. For a circumprimary disc, the location of the outer edge is determined by tides and an approximate value for its location is given by $r_{\rm t} \simeq 0.9 r_{\rm L1}$ \citep{Frank:2002aa}. Here, $r_{\rm L1}$ is the primary's Roche lobe radius which is given by \citep{Eggleton:1983aa}
\begin{equation}
\frac{r_{L1}}{r_{\rm b}} \simeq \frac{0.49(M_1/M_2)^{-2/3}}{0.6(M_1/M_2)^{-2/3}+\ln [1+(M_1/M_2)^{-1/3}]}.
\end{equation}
A more accurate method to find the tidal truncation radius is described, and the radii outside which tidal torques dominate are tabulated, by \cite{Papaloizou:1977aa} for various mass ratios (see their Table 1). These values are a few per cent smaller than those estimated from the Roche-lobe approximation. However, for our purpose an approximate value will suffice and we overlay (dotted line) the approximate location of the outer disc for different binary mass ratios in Fig.~\ref{fig:cp-cb} (left-hand panel). As we see in Fig. \ref{fig:cp-cb} (left-hand panel), the deviation from Keplerian rotation remains small for all parameters and is maximized for larger mass companions. For circumprimary discs the dimensionless orbital shear lies in the range of $1.5 \leq q \leq 1.55$, and correspondingly $1 \leq \tilde{\kappa}^2 \leq 0.9$.

\subsubsection{Circumbinary discs}
\label{sec:nonkep3}
Similarly to the previous case, we can derive the orbital and epicyclic frequencies for a circumbinary disc from the gravitational potential of the binary. Following, e.g., \cite{Nixon:2011ab}, we assume that the relevant potential is given by the $m=0$ Fourier mode of the time-dependent potential generated by the binary orbit, as all other modes are oscillatory with time and thus their effects cancel out over a binary period. Thus we can write the potential due to the binary as the sum of the potentials from each binary component averaged over their orbits (which for simplicity we assume here to be circular). This means spreading the mass of each binary component ($M_1$, $M_2$) into rings at their orbital radii from the origin ($a_1$, $a_2$). Note that the total mass $M=M_1+M_2$ and the orbital separation $r_{\rm b} = a_1+a_2$, and thus $a_1 = r_{\rm b}M_2/M$ and $a_2 = r_{\rm b}M_1/M$. For an arbitrary position in cylindrical polars $(r,z)$ with the binary in the $z=0$ plane (note that by symmetry we are free to take the arbitrary position to have any azimuthal angle, and thus we take it to be zero), the $m=0$ mode of the gravitational potential is given by
\begin{equation}
  \Phi = \Phi_1 + \Phi_2 = \frac{GM_1}{2\pi}\int_0^{2\pi}\frac{d\phi}{r_1} + \frac{GM_2}{2\pi}\int_0^{2\pi}\frac{d\phi}{r_2}\,, 
\end{equation}
where $r_1$ is the distance to the $M_1$ ring from the point $(r,z)$ with $r_1^2 = r^2 + a_1^2 + z^2 - 2ra_1\cos\phi$, and $r_2$ is the distance to the $M_2$ ring from the point $(r,z)$ with $r_2^2 = r^2 + a_2^2 + z^2 - 2ra_2\cos\phi$, and $\phi$ is the azimuthal angle around the rings. The orbital and epicyclic frequencies can now be derived from the potential using the definition for the orbital frequency
\begin{equation}
  \Omega^2 = \left.\frac{1}{r}\frac{\partial \Phi}{\partial r}\right|_{z=0}\,,
\end{equation}
and equation~\ref{eq:kappa} above for the epicyclic frequency (again evaluated at $z=0$).

In the limit that $r/r_{\rm b}\gg 1$, analytical expressions for the frequencies are given by e.g. \cite{Nixon:2011aa,Facchini:2013aa,Miranda:2015aa}. Here, we are concerned with small radii near the binary, and so we choose to solve the integrals numerically to avoid taking the large radius approximation. For $r/r_{\rm b} \gg 1$, the results are essentially identical to the approximate expressions. For $r\sim r_{\rm b}$, the frequencies differ from the approximate expressions by several 10s of per cent. To determine the largest deviation from Keplerian rotation in this case, we need to determine the location of the inner disc edge. To illustrate the inner disc location, we use tidal truncation radii given by \cite{Artymowicz:1994aa} (see their Table 1) for a circumbinary disc around a circular binary with various mass ratios.

In Fig. \ref{fig:cp-cb} (right-hand panel), we show the dimensionless shear rates and corresponding epicyclic frequencies for the circumbinary discs as a function of $r/r_{\rm b}$ for various mass ratios. The effect of non-Keplerian rotation is maximum when the binary components have equal masses. In this case, the dimensionless orbital shear and dimensionless epicyclic frequency lie in the range $1.5 \leq q \leq 1.57$ and $1 \leq \tilde{\kappa}^2 \leq 0.86$. The strength of the deviation from Keplerian rotation in circumbinary discs is similar to the circumstellar disc case, with both remaining much closer to Keplerian than the inner regions of discs around black holes (see below).

\subsection{Discs around Kerr black holes}
\label{sec:nonkep1}
\begin{figure}
  \begin{center}
        {\includegraphics[angle=270,scale=.43]{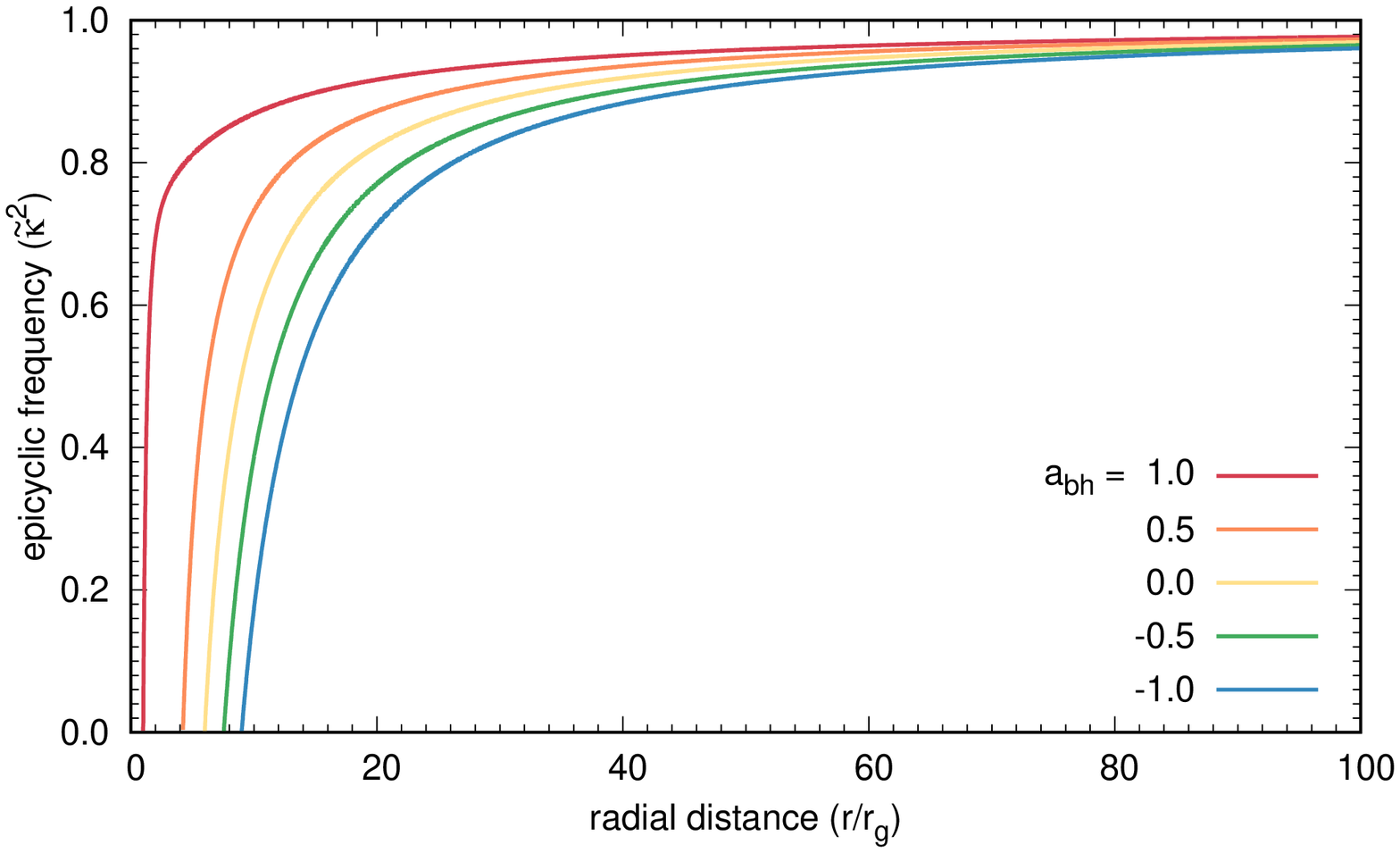}}
        {\includegraphics[angle=270,scale=.43]{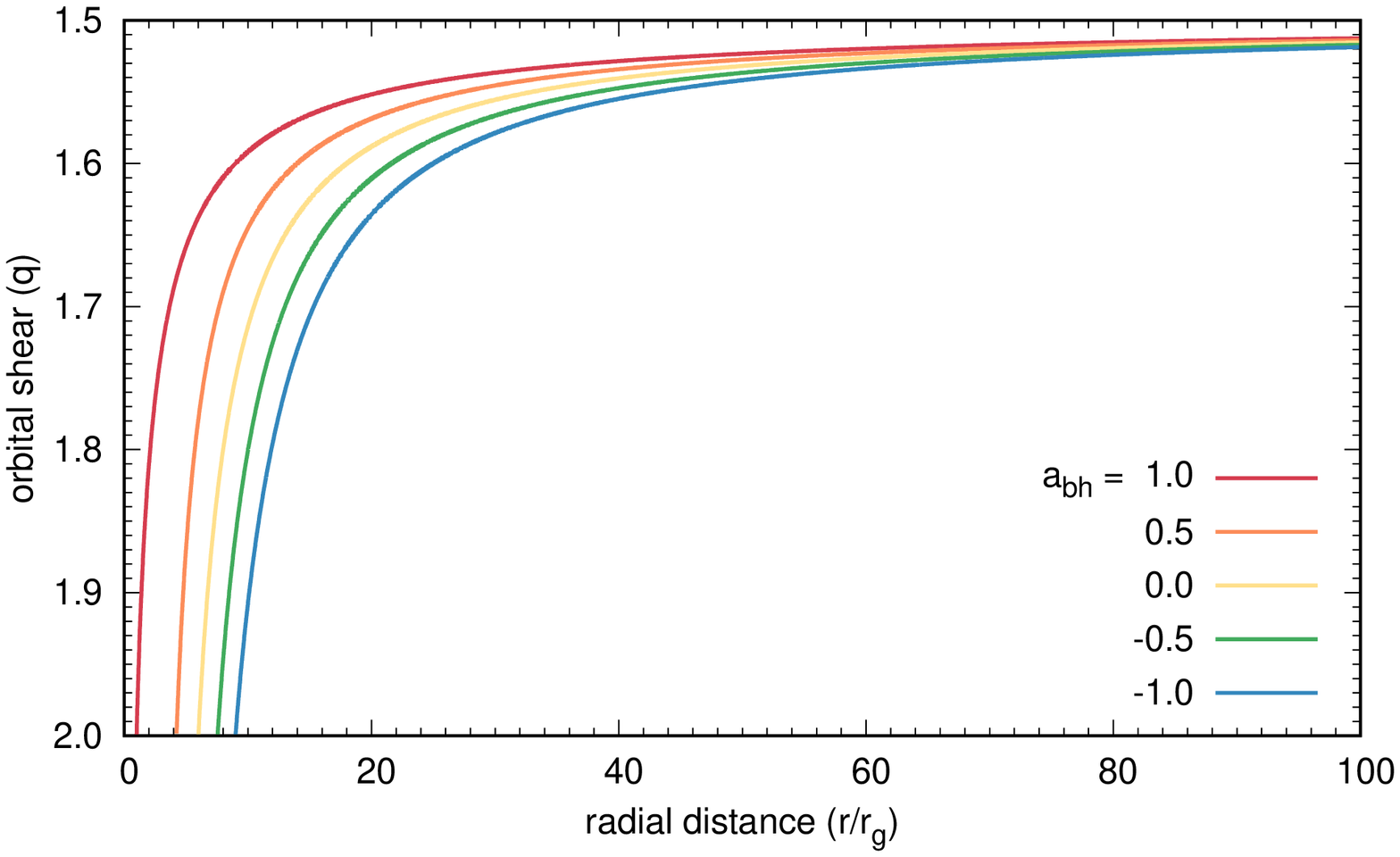}}
              \caption{We show the radial distributions of the dimensionless epicyclic frequency from equation~\ref{eq:kappa_bh} (left-hand panel) and the dimensionless shear rate from equation~\ref{eq:q_bh} (right-hand panel) for discs around spinning black holes for various spin parameters. The inner radii for each case corresponds to the ISCO where $\tilde{\kappa}^2_{\rm bh} = 0$ and $q_{\rm bh}=2$. The deviation from Keplerian rotation is large in the inner parts of the disc with $\tilde{\kappa}^2_{\rm bh} \rightarrow 0$, and correspondingly $q_{\rm bh} \rightarrow 2$. Retrograde discs are more strongly affected at the same radius than prograde discs.}
        \label{fig:bh}
  \end{center}
\end{figure}

For discs around black holes the orbital frequencies deviate from Keplerian values leading to apsidal precession and (for misaligned orbits around spinning black holes) nodal precession. These effects are strongest at small radii around the black hole. The stable orbit of accreting gas in a thin disc around a spinning black hole can be described by a circular equatorial Kerr geodesic with orbital frequency given by \citep{Bardeen:1972,Kato:1990aa,Gammie:2004}
\begin{equation}
\label{omega_bh}
\Omega^{-1}=\left(\frac{GM}{c^3}\right)\left[\left(r/r_{\rm g}\right)^{3/2}+a_{\rm bh}\right]\,.
\end{equation}
Here, $r$ is the distance from the black hole in Boyer-Lindquist coordinates, $r_{\rm g}=GM/c^2$ is the gravitational radius of the black hole, $a_{\rm bh}$ is the dimensionless spin parameter of the black hole with $-1 \leq a_{\rm bh} \leq 1$. A non-spinning black hole has $a_{\rm bh}=0$, and $a_{\rm bh}<0$ implies a retrograde disc. The radial dependence of the epicyclic frequency for circular equatorial Kerr geodesics is given by \citep{Gammie:2004}
\begin{equation}
\label{eq:kappa_bh}
\kappa^2_{\rm bh}=\frac{1}{r^3}\frac{1-6(r/r_g)^{-1}+8a(r/r_g)^{-3/2}-3a^2(r/r_g)^{-2}}{1-3(r/r_g)^{-1}+2a(r/r_g)^{-3/2}}\,,  
\end{equation}
in which $G=M=c=1$. The epicyclic frequency is zero ($\kappa^2_{\rm bh} = 0$) at the innermost stable circular orbit (ISCO), and imaginary inside the ISCO ($\kappa^2_{\rm bh} < 0$) where a circular shear flow around the hole is unstable to axisymmetric perturbations. As mentioned above, in a Keplerian potential the epicyclic frequency is equal to the Keplerian orbital frequency, i.e. $\tilde{\kappa}^2 = 1$. However, near a black hole the epicyclic frequency is less than the orbital frequency. A thin disc starts with $ \tilde{\kappa}^2_{\rm bh} = 0$ at the ISCO, and then $\tilde{\kappa}^2_{\rm bh}$ increases until it reaches 1 at large radii. Fig. \ref{fig:bh} (left-hand panel) shows the radial distribution of the dimensionless epicyclic frequency, $\tilde{\kappa}^2_{\rm bh}$, for various $a_{\rm bh}$ values. The deviation from Keplerian rotation is stronger (weaker) when the disc is retrograde (prograde), and maximum deviation at a given radius occurs when $a_{\rm bh}=-1$.

A dimensionless measure of the shear rate for circular equatorial geodesics in the Kerr metric is given by \citep[][see also \citealt{Penna:2013}]{Gammie:2004}
\begin{equation}
\label{eq:q_bh}
q_{\rm bh}=\frac{3}{2}\frac{1-2(r/r_g)^{-1}+a^2(r/r_g)^{-2}}{1-3(r/r_g)^{-1}+2a(r/r_g)^{-3/2}}.
\end{equation}
Circular equatorial geodesics have $q_{\rm bh} = 2$ at the ISCO, and $q_{\rm bh}$ becomes 3/2 at large radii. Fig. \ref{fig:bh} (right-hand panel) shows the radial distribution of the dimensionless shear rate for circular equatorial geodesics around a spinning black hole. We should note that Fig. \ref{fig:bh} shows the epicyclic frequency and the shear rate values for the region outside the ISCO. The inner disc radii for each $a_{\rm bh}$ value can be found from the relation between the black hole spin parameter and the location of the ISCO \citep[e.g.][]{King:2006aa}:
\begin{equation}
\label{eq:isco}
a_{\rm bh}=\frac{1}{3}\big(r_{\rm ISCO}/r_{\rm g}\big)^{1/2}\Big[4-\Big(3\big(r_{\rm ISCO}/r_{\rm g}\big)-2\Big)^{1/2}\Big]
\end{equation}
These locations correspond to the radii where $\tilde{\kappa}^2_{\rm bh}$ changes sign.

We should also note that the relativistic expressions of the dimensionless epicyclic frequency ($\kappa_{\rm bh}$) and the dimensionless shear rate ($q_{\rm bh}$) given by (\ref{eq:kappa_bh}) and (\ref{eq:q_bh}) do not exactly satisfy the simple relation given by (\ref{eq:kappa_q}). Here, the relation between $\kappa_{\rm bh}$ and $q_{\rm bh}$ is modified as follows:
\begin{equation}
\label{eq:kappa_q_bh}
\tilde{\kappa}^2_{\rm bh}=\big(4-2q_{\rm bh}\big)\big[1+a_{\rm bh}(r/r_g)^{-3/2}\big]^2.
\end{equation}
We see that (\ref{eq:kappa_q_bh}) is reduced to (\ref{eq:kappa_q}) at large radii where $\big[1+a_{\rm bh}(r/r_g)^{-3/2}\big] \rightarrow 1$, or in case of a Schwarzschild black hole, i.e. where $a_{\rm bh} = 0$. We comment on this further below in the context of our results.\footnote{We also note that the frequencies used here differ from those used in \cite{Lubow:2002aa} which were derived in the frame of an observer at infinity by \cite{Kato:1990aa}. We use the frequencies derived in the fluid frame by \cite{Gammie:2004}, as the calculation of the torque coefficients below (which depends on the shear rate $q$) is a local fluid calculation and the method we use (following \citealt{Ogilvie:2013aa}) requires that the relation in (\ref{eq:kappa_q}) holds.}
 
\section{Instability of a non-Keplerian disc}
\label{sec:analysis}
The local stability of an isolated, Keplerian, warped disc with $\alpha > H/R$ (where $\alpha$ is the dimensionless \citealt{Shakura:1973aa} viscosity parameter and $H/R$ is the disc angular semi-thickness) has been described in \cite{Ogilvie:2000aa} and \cite{Dogan:2018aa}. In this work we follow the same analysis, but focus on the effect of non-Keplerian rotation. Here we provide a brief description of the relevant equations and refer the reader to \cite{Ogilvie:2000aa} and \cite{Dogan:2018aa} for more details. The governing evolutionary equations are the conservation of mass equation,
\begin{equation}
 \label{eq:conservation1}
\frac{\partial \Sigma}{\partial t}+\frac{1}{r}\frac{\partial}{\partial r}(r\bar{\upsilon_r}\Sigma)=0\,,
\end{equation}
and the conservation of angular momentum equation,
\begin{equation}
  \label{eq:here}
  \frac{\partial }{\partial t}\left(\Sigma r^2\Omega \mathbfit{l}\right)
  =\frac{1}{r}\frac{\partial}{\partial r}\left[Q_1\Sigma c_{\rm s}^2 r^2\mathbfit{l} + Q_2\Sigma c_{\rm s}^2 r^3\frac{\partial \mathbfit{l}}{\partial r} + Q_3\Sigma c_{\rm s}^2 r^3\mathbfit{l}\times\frac{\partial \mathbfit{l}}{\partial r} - \left(\frac{\partial}{\partial r}\left[Q_1\Sigma c_{\rm s}^2 r^2\right] - Q_2\Sigma c_{\rm s}^2r\left|\psi\right|^2\right)\frac{h}{h^\prime}\mathbfit{l}\right]\,.
\end{equation}
Here $\Sigma\left(R,t\right)$ is the disc surface density, $\bar{\upsilon}_r$ is the mean radial velocity, $\Omega\left(R\right)$ is the orbital angular velocity of each annulus of the disc, $\mathbfit{l}\left(R,t\right)$ is the unit angular momentum vector pointing perpendicular to the local orbital plane, $c_{\rm s}$ is the sound speed, $\left|\psi\right| = r|\partial \mathbfit{l}/\partial r|$ is the warp amplitude, $h = r^2\Omega$ is the specific angular momentum, $h^\prime = {\rm d} h/{\rm d} r$ and $Q_i$ are the dimensionless torque coefficients. A detailed description of these equations can be found in \cite{Ogilvie:2000aa,Ogilvie:2013aa,Dogan:2018aa}. Recall that $q = -{\rm d}\ln \Omega/{\rm d}\ln r$, and here we are focusing on non-Keplerian disc rotation, implying that $Q_i = Q_i\left(\alpha,q,\left|\psi\right|\right)$. Note we have assumed no bulk viscosity ($\alpha_{\rm b} = 0$).

The details of the stability analysis are given in \citep{Ogilvie:2000aa} and \cite{Dogan:2018aa}. The stability is considered with respect to linear perturbations in $\delta\Sigma$ and $\delta\mathbfit{l}$, and then wave solutions are sought. This yields a third order dispersion relation given by
\begin{equation}
\label{eq:dr}
\setlength\jot{12pt}
\begin{split}
    s^3&-s^2\left[aQ_1-2Q_2+|\psi|\Big(aQ'_1-Q'_2\Big)\right]-s\left[2a{Q}_1{Q}_2-Q_2^2-Q_3^2+|\psi|\Big(a{Q}_1{Q}'_2-Q_2Q'_2-Q_3Q'_3\Big)\right]\\
   &-a\left[Q_1(Q_2^2+Q_3^2)+|\psi|\Big(Q_1Q_2Q'_2-Q_1'Q_2^2+Q_1Q_3Q'_3-Q_1'Q_3^2\Big)
   \right]=0.
\end{split}
\end{equation}
Here, the prime on $Q_i$ represents differentiation with respect to $|\psi|$, $a = h/rh^\prime = {\rm d}\ln r/{\rm d}\ln h = 1/(2-q)$. We note that $a=2$ for a Keplerian disc with $q=3/2$. The dimensionless growth rate, $s$, is defined by
\begin{equation}
  s=-\frac{{\rm i}\omega}{\Omega}\Bigg(\frac{\Omega}{c_s k}\Bigg)^2
\end{equation}
where $k \lesssim 1/H$ for validity. The perturbations grow ($\Re[s]>0$) or decay ($\Re[s]<0$) as $\exp[\Re(-i\omega)t]$.
\begin{figure}
  \begin{center}
  \begin{tabular}{ccc}
  $\alpha = 0.01$ & $\alpha = 0.03$ & $\alpha = 0.10$\\
         {\includegraphics[angle=270,scale=.43]{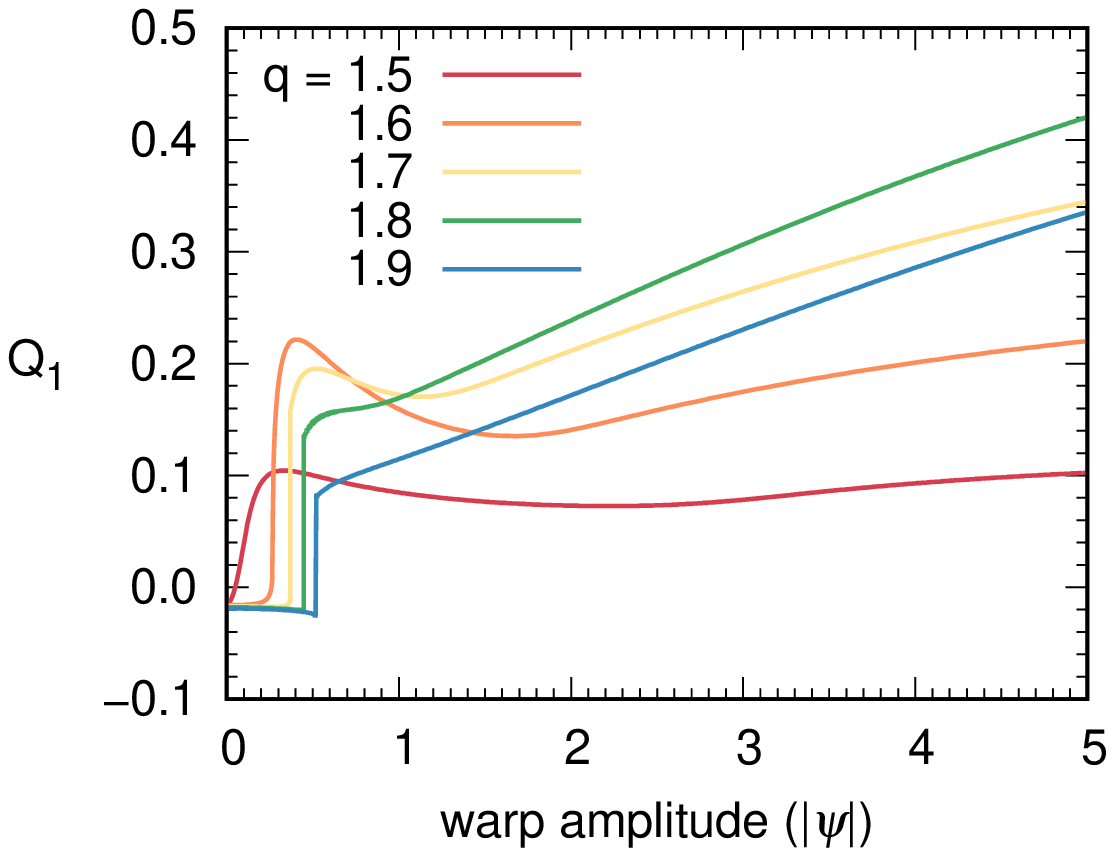}}&
         {\includegraphics[angle=270,scale=.43]{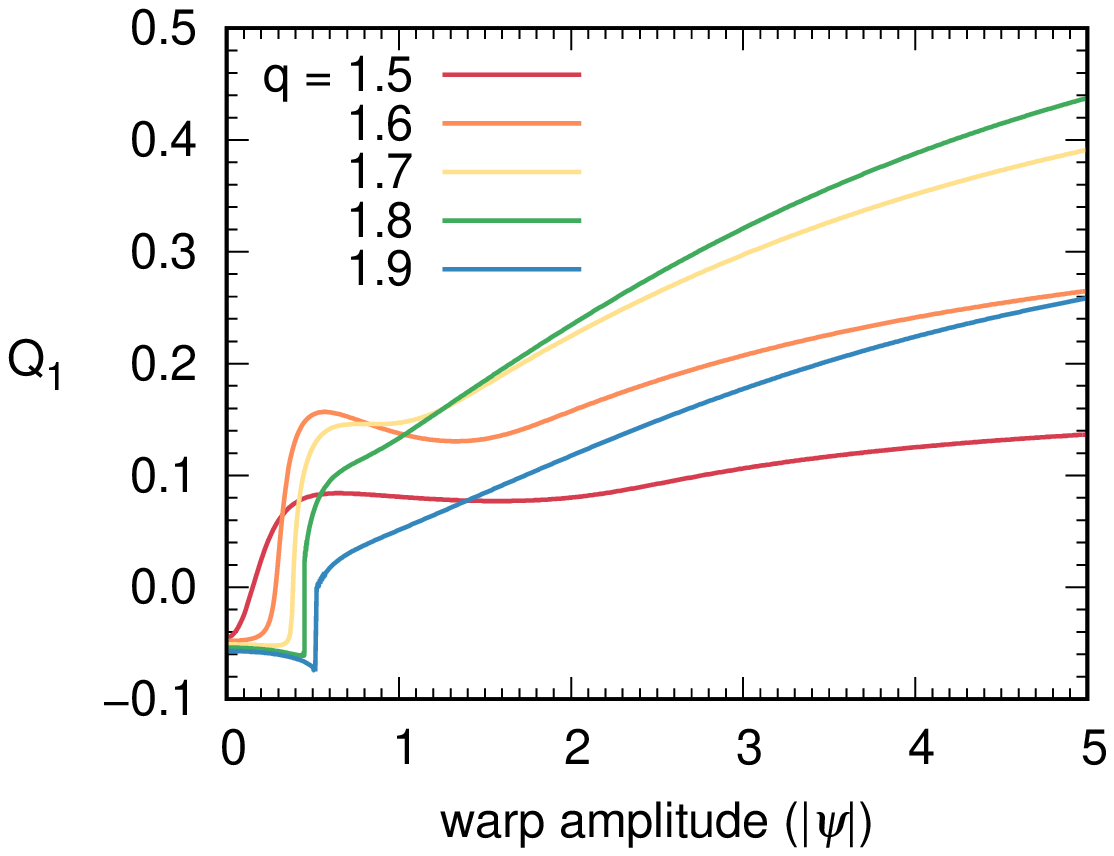}}&
         {\includegraphics[angle=270,scale=.43]{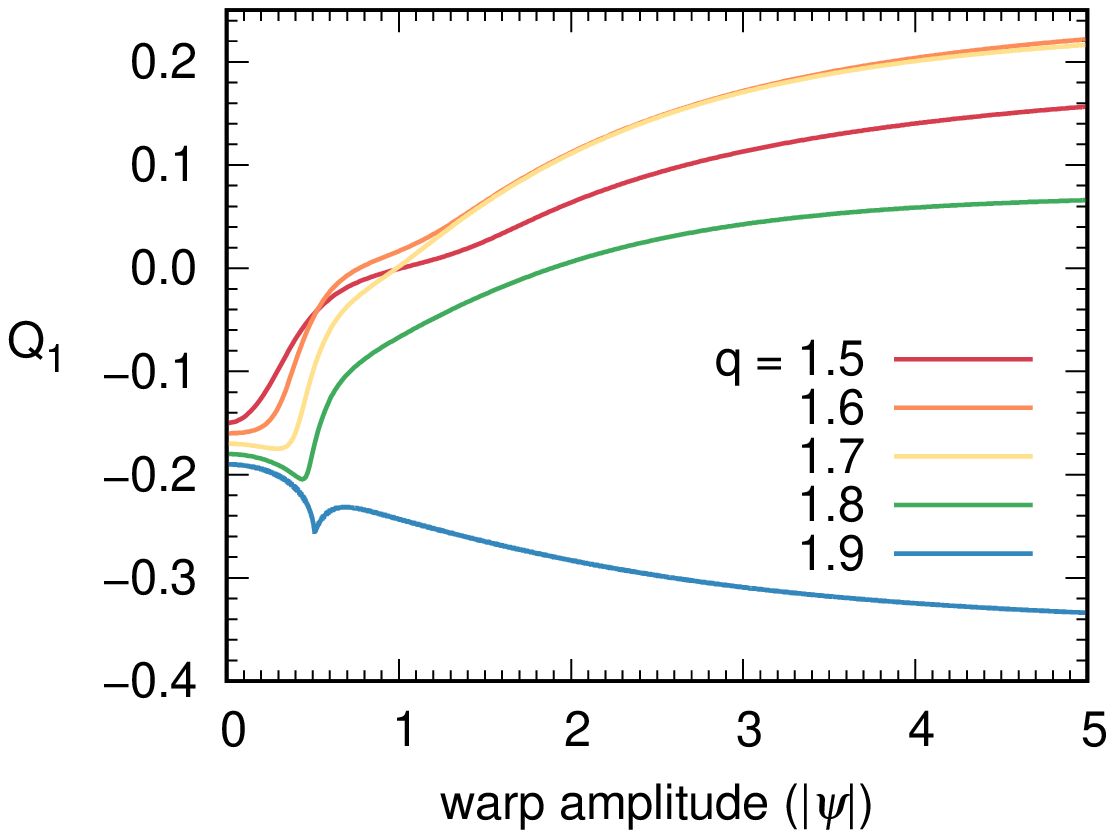}}\\
         {\includegraphics[angle=270,scale=.43]{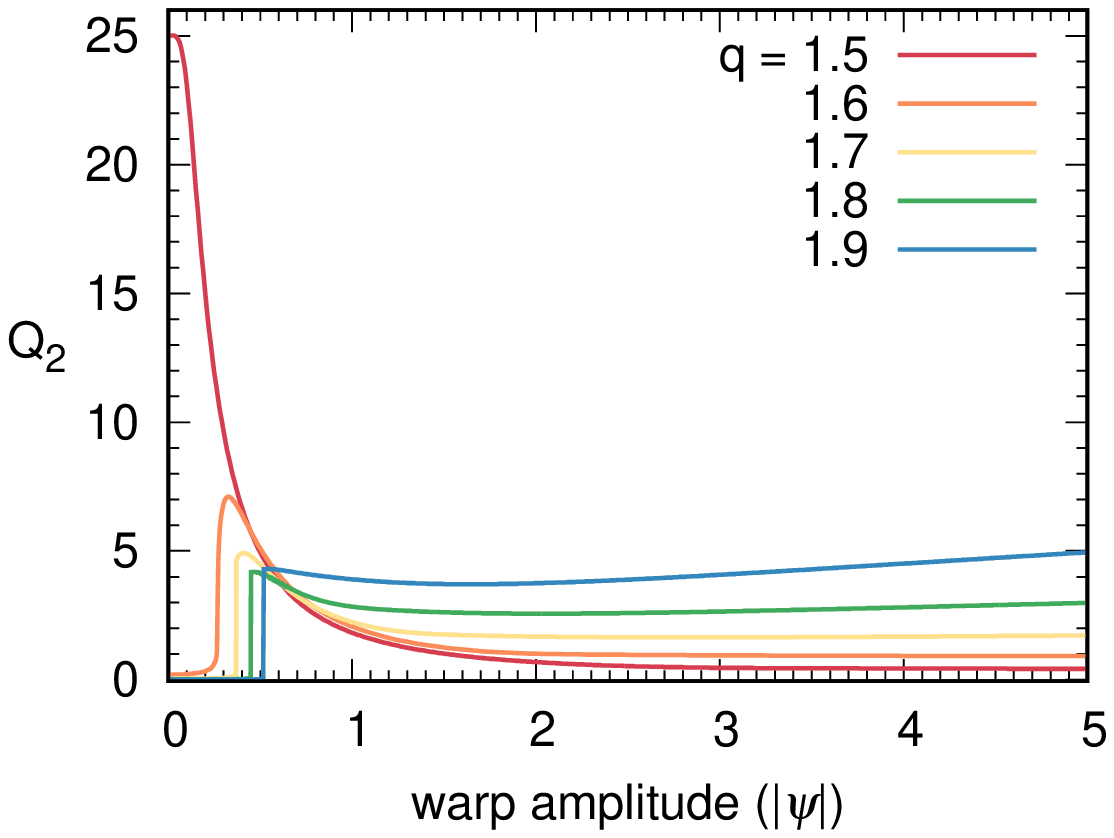}}&
         {\includegraphics[angle=270,scale=.43]{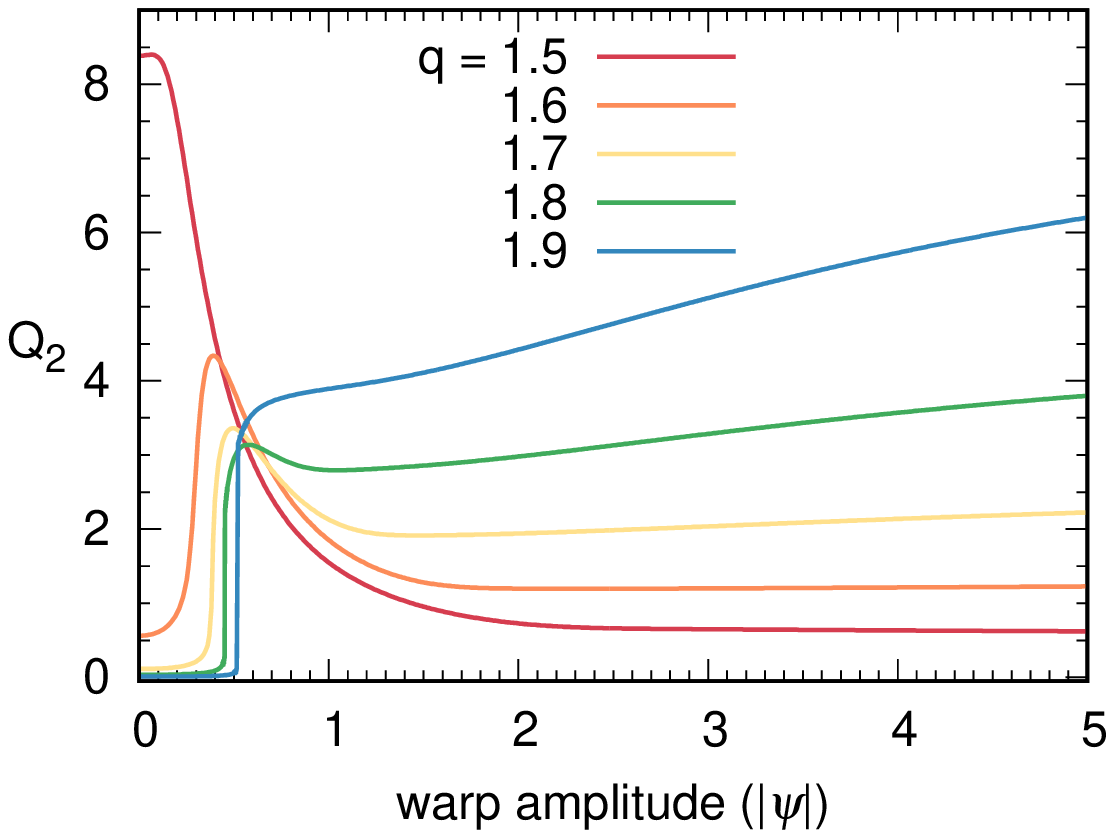}}&
         {\includegraphics[angle=270,scale=.43]{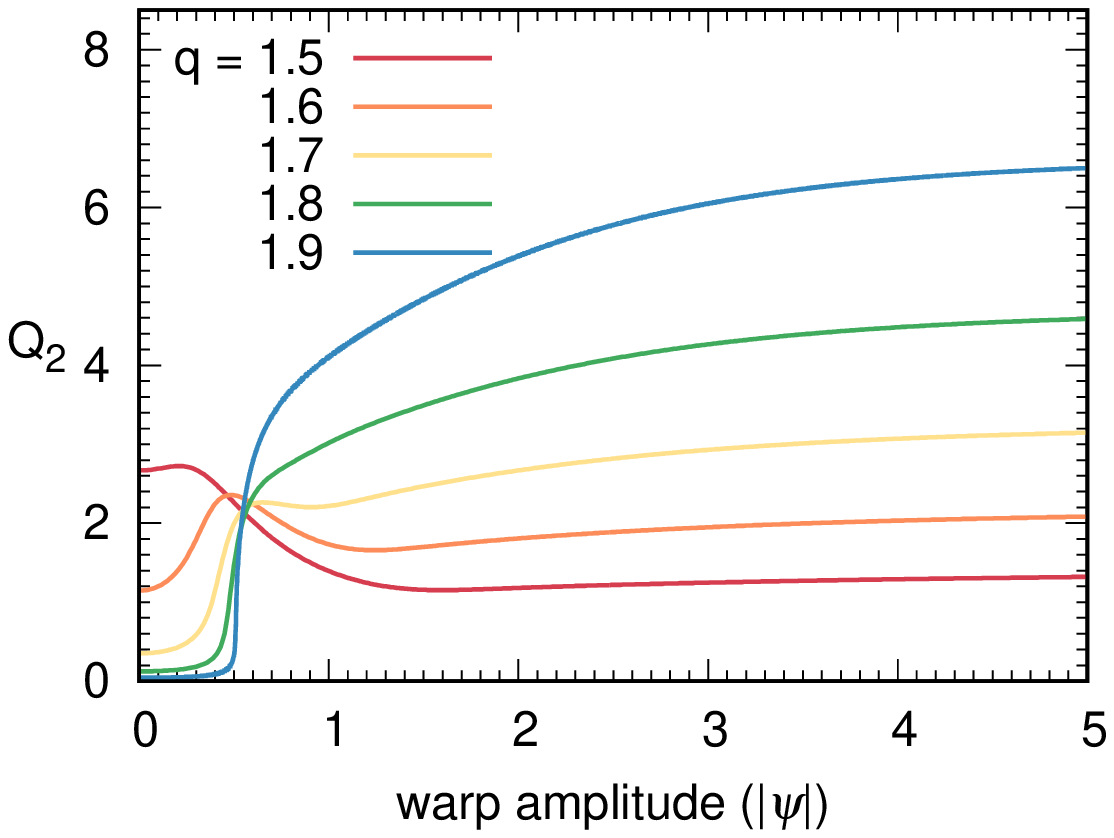}}\\
         {\includegraphics[angle=270,scale=.43]{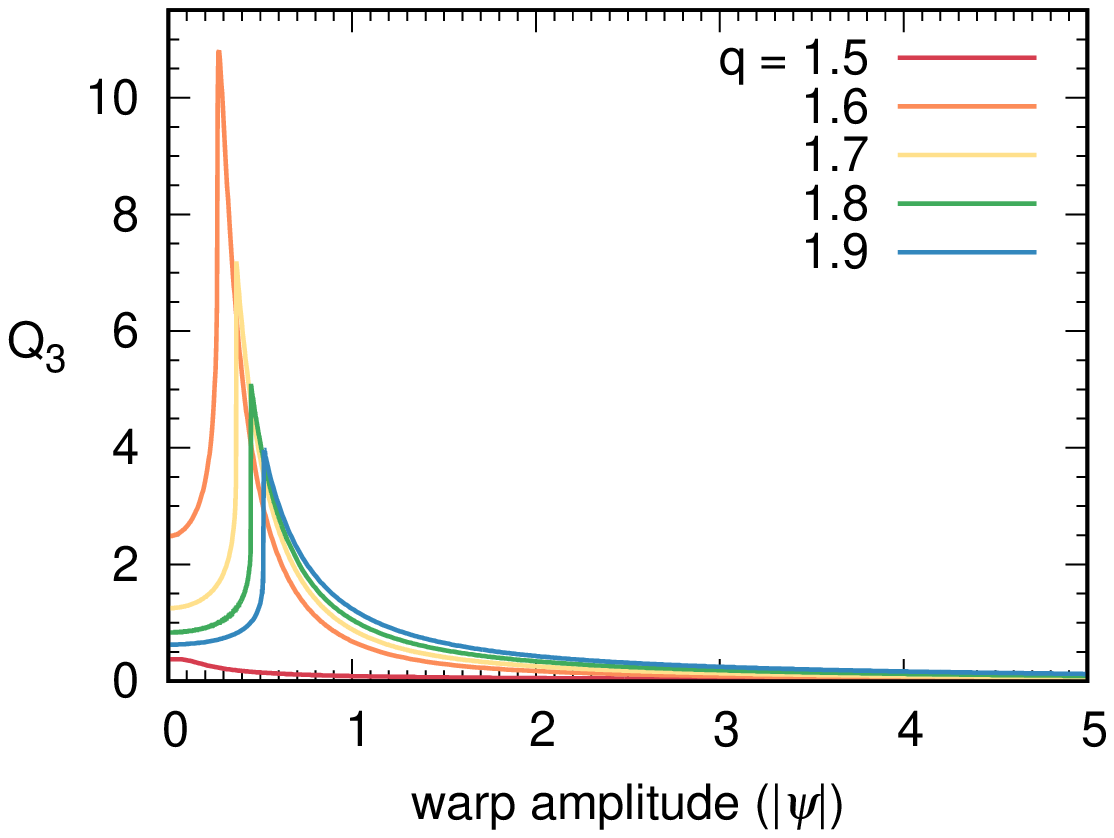}}&
         {\includegraphics[angle=270,scale=.43]{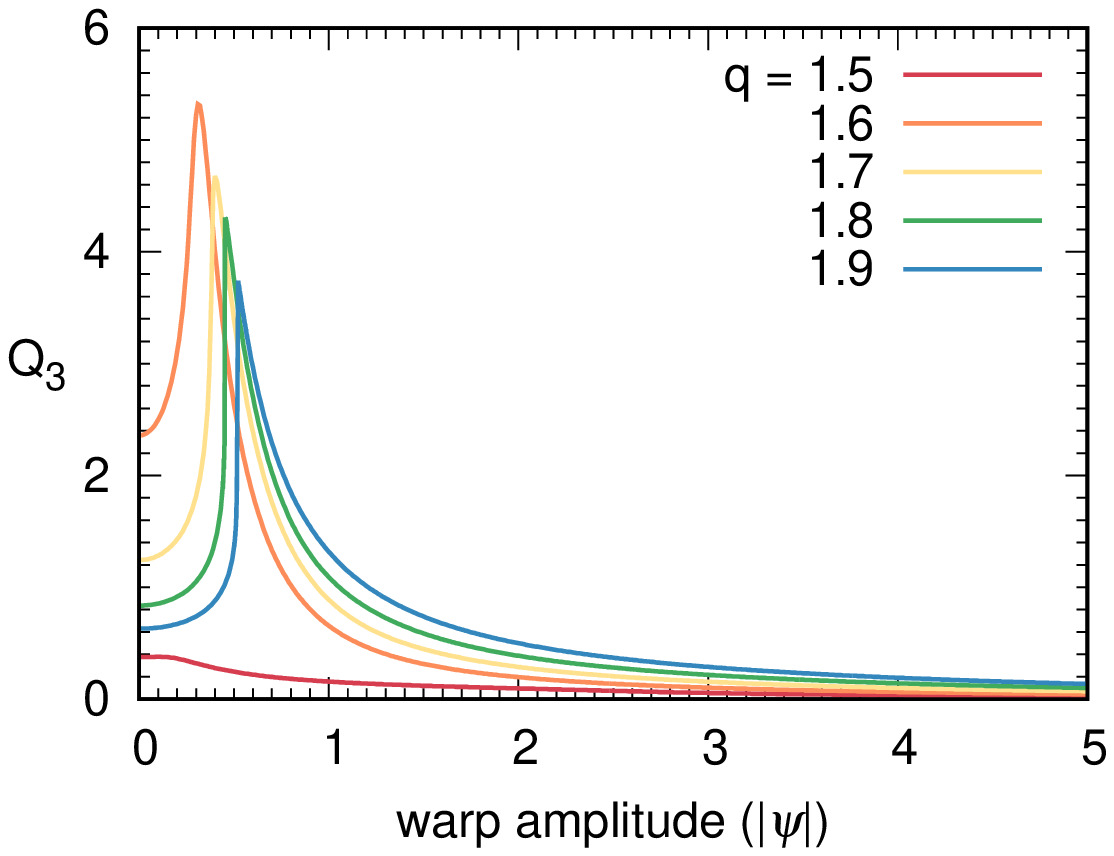}}&
         {\includegraphics[angle=270,scale=.43]{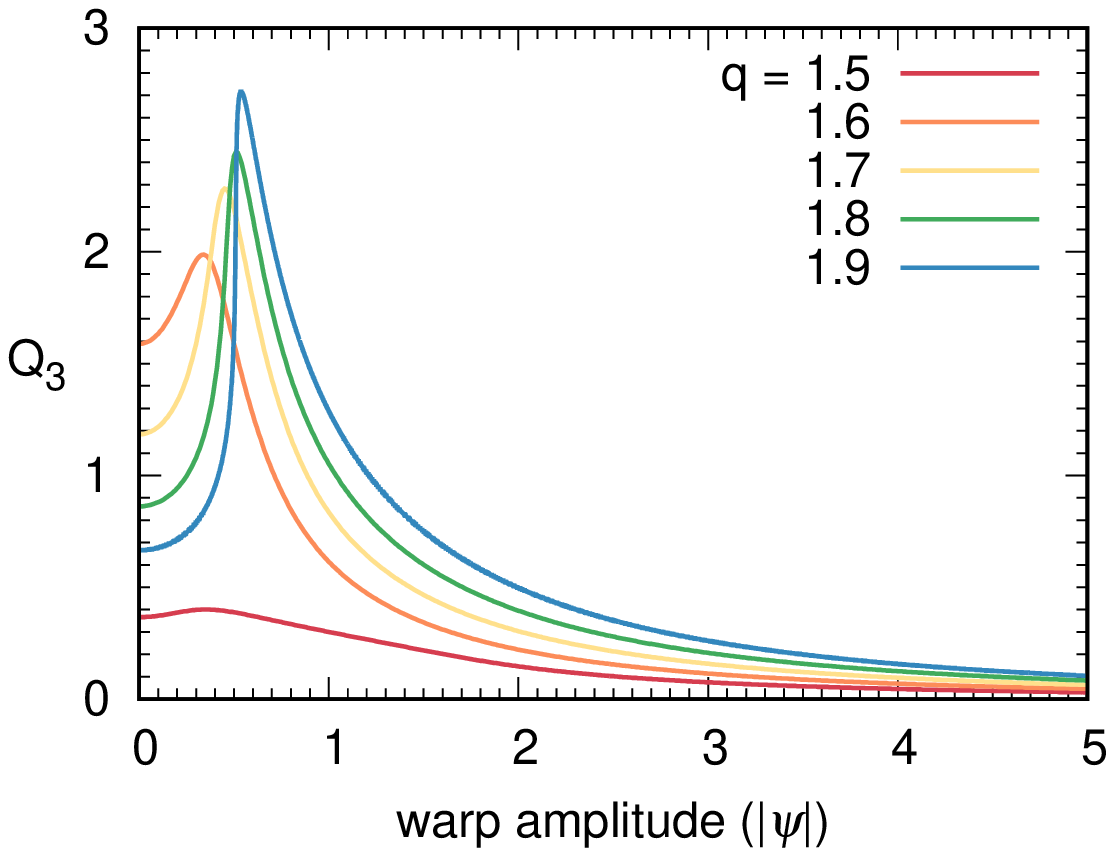}}
         \end{tabular}
              \caption{$Q_1$ (top), $Q_2$ (middle) and $Q_3$ (bottom) coefficients plotted as a function of warp amplitude $|\psi|$, assuming that $\alpha = 0.01$ (first column), $\alpha = 0.03$ (second column), $\alpha = 0.1$ (third column) for various $q$ values. The red line represents the Keplerian case in each panel, i.e. $q = 1.5$.}
        \label{fig:Qi}
  \end{center}
\end{figure}

\cite{Dogan:2018aa} provide a detailed investigation with simplified cases as well as full solutions to (\ref{eq:dr}). The disc becomes unstable if any of the roots of (\ref{eq:dr}) has a positive real part, i.e. $\Re(s)>0$, as the perturbations then grow exponentially with time. In \cite{Dogan:2018aa}, we showed that there is always a critical warp amplitude, $|\psi|_c$, which gives rise to instability for the parameters we have explored, with the exception of high viscosity discs with $\alpha \sim 1$. Nearly flat discs with $|\psi| \lesssim 0.1$ were found to be stable. The instability occurs as a result of anti-diffusion of the warp amplitude. The growth rates of the instability can be comparable with the dynamical rate ($\Re\left[s\right] \sim 1$).

The main effect of non-Keplerian rotation is to change the values of the dimensionless torque coefficients in the conservation of angular momentum equation (\ref{eq:here}). Other than that, the basic equations remain the same. Similar to our previous work, we calculate the $Q_i$ coefficients using the code kindly provided by G.~I.~Ogilvie following the calculations in \cite{Ogilvie:2013aa}. Fig. \ref{fig:Qi} shows the values of $Q_i$ coefficients for three different values of the shear viscosity parameter $\alpha=0.01, 0.03$ and $0.1$, for various $q$ values. The $Q_i$ coefficients are calculated assuming a locally isothermal equation of state, and therefore relevant to optically thin discs. However, as discussed in \cite{Dogan:2018aa}, our results are also applicable to the highly optically thick case as the basic physics in these two cases remains the same on the timescale on which the torque coefficients, and thus the stability, is determined. We note that some part of the solutions of the $Q_i$ coefficients for non-Keplerian discs show a discontinuous behaviour in the regions where they change their sign from negative to positive. This occurs when the viscosity is very low. This behaviour was discussed by \cite{Ogilvie:1999aa} and \cite{Ogilvie:2013aa} for inviscid ($\alpha= 0$) and non-Keplerian discs with $q > 1.5$ ($\tilde{\kappa}^2 < 1$). The breakdown of the solution was attributed to a non-linear resonance of coupled horizontal and vertical oscillators (see Appendix B of \citealt{Ogilvie:2013aa}). When $\alpha=0.01$, we observe a similar behaviour for $q > 1.65$, and similarly when $\alpha= 0.03$, we observe this behaviour for $q > 1.75$. When the viscosity is higher (e.g. for $\alpha=0.1$), this behaviour disappears.

\begin{figure}
  \begin{center}
    {\includegraphics[angle=270,scale=.46]{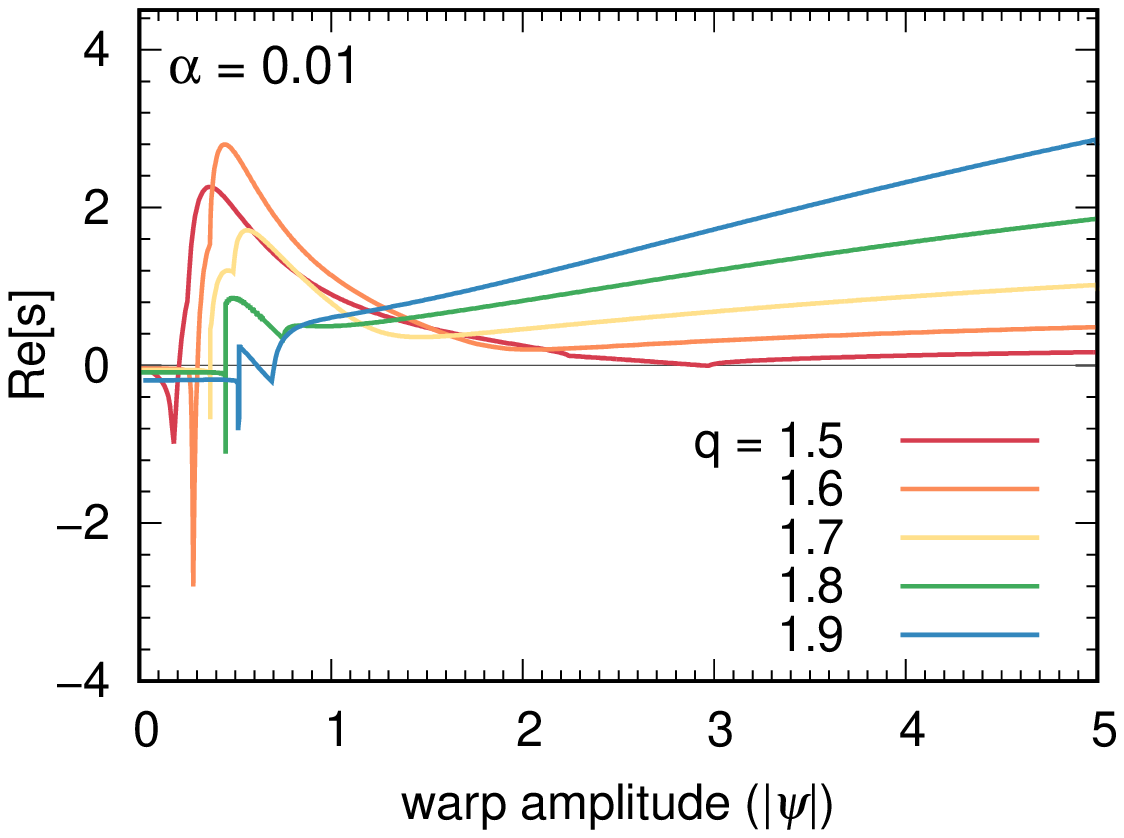}}
    {\includegraphics[angle=270,scale=.46]{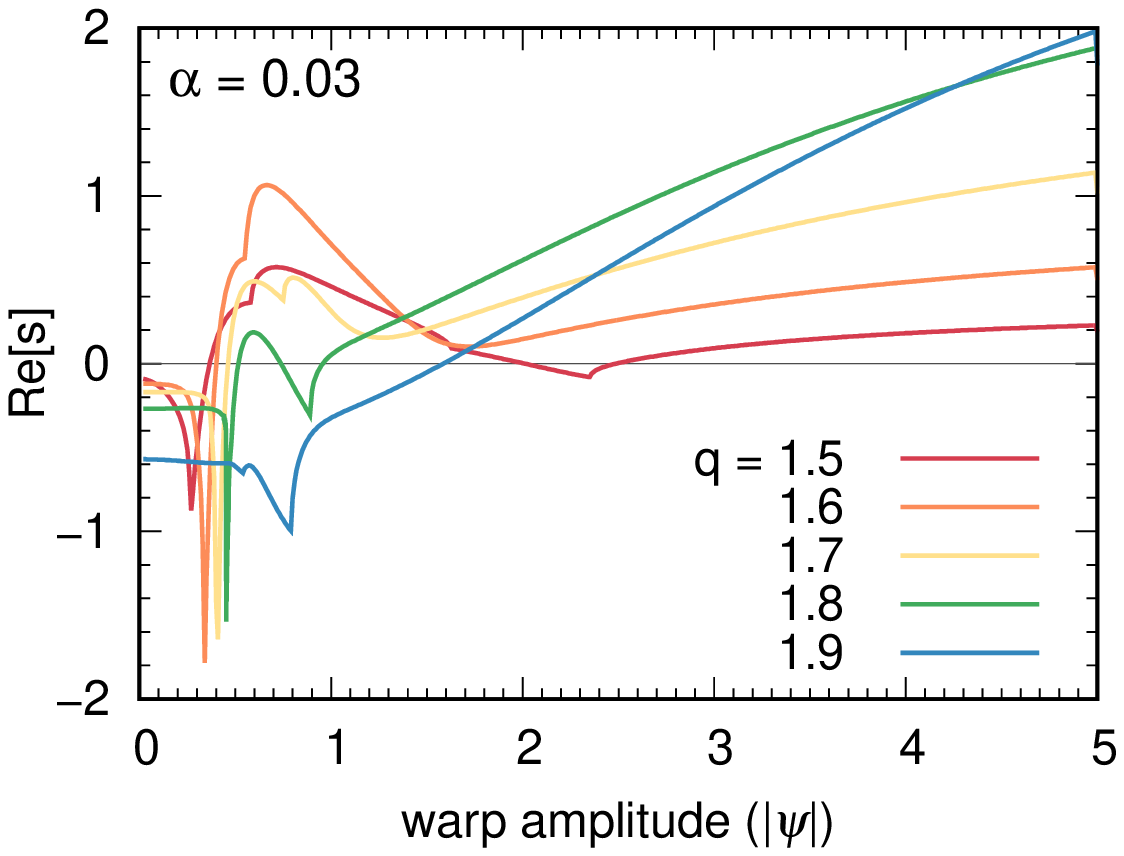}}
    {\includegraphics[angle=270,scale=.46]{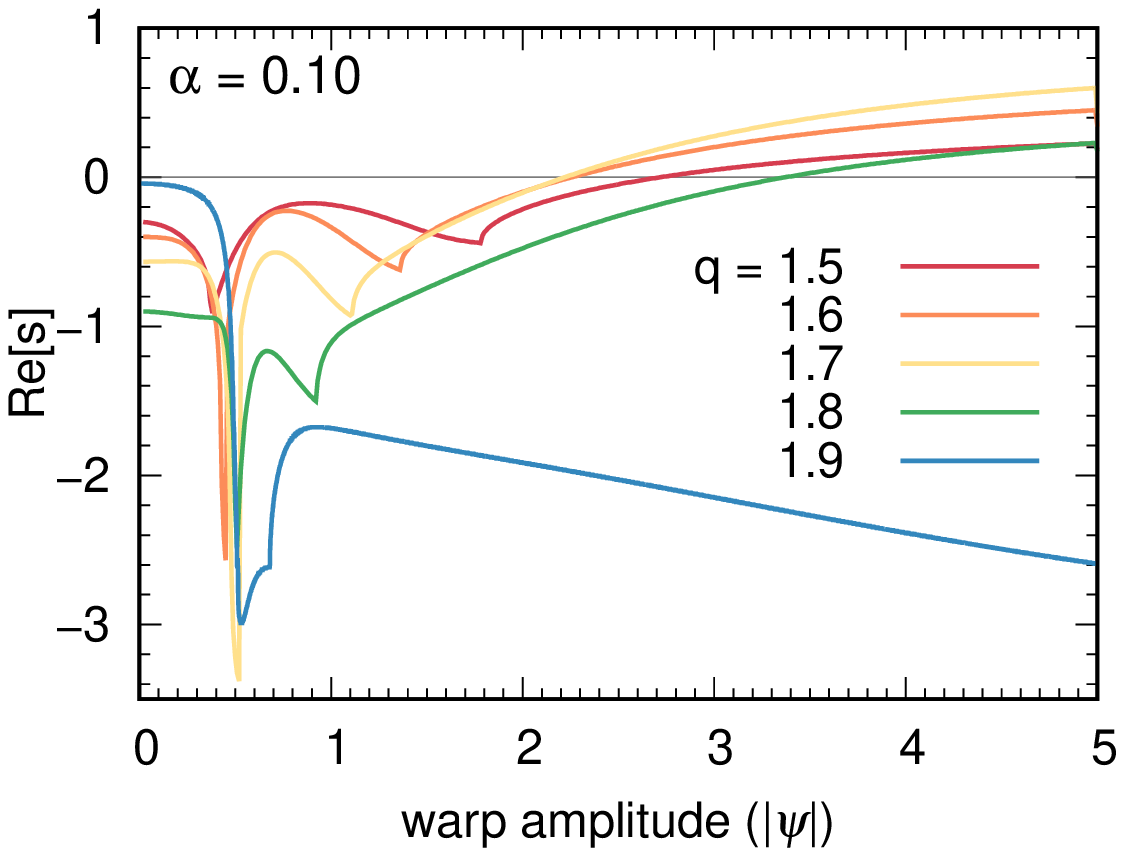}}
    \caption{We show the dimensionless growth rates $\mathfrak{R}[s]$ as functions of $|\psi|$ for various values of $q$ for a disc with $\alpha = 0.01$, $\alpha = 0.03$ and $\alpha = 0.1$. The growth rates are calculated by solving equation~\ref{eq:dr} \citep[the full solutions are given in][their eqns~$33-37$]{Dogan:2018aa}. The red line represents the Keplerian case where $q = 3/2$ and the grey line represents the zero growth rate in each panel.}
    \label{fig:growthrates}
  \end{center}
\end{figure}

By using the non-Keplerian values of the $Q_i$ coefficients, we evaluate the growth rates of the instability by solving equation~\ref{eq:dr} \citep[the full solutions are given in][their eqns~$33-37$]{Dogan:2018aa}. In Fig.~\ref{fig:growthrates}, we show the variation of dimensionless growth rates $\mathfrak{R}[s]$ with $|\psi|$ for different values of $q$. The figures are plotted for $\alpha = 0.01$, $\alpha = 0.03$ and $\alpha = 0.1$. We note that these growth rates mainly consist of the combined effects of the $Q_1$ and $Q_2$ torque coefficients (see Fig. 2 in \citealt{Dogan:2018aa}). At small warp amplitudes the $Q_2$ term is dominant. At larger values of the warp amplitude, the $Q_1$ term becomes dominant. In Fig. \ref{fig:growthrates} we see that the growth rates of the instability are higher and the critical warp amplitudes, where the disc becomes unstable, are smaller for low $\alpha$ as found for the Keplerian case in \cite{Dogan:2018aa}.

\begin{figure}
  \begin{center}
    {\includegraphics[angle=270,scale=.55]{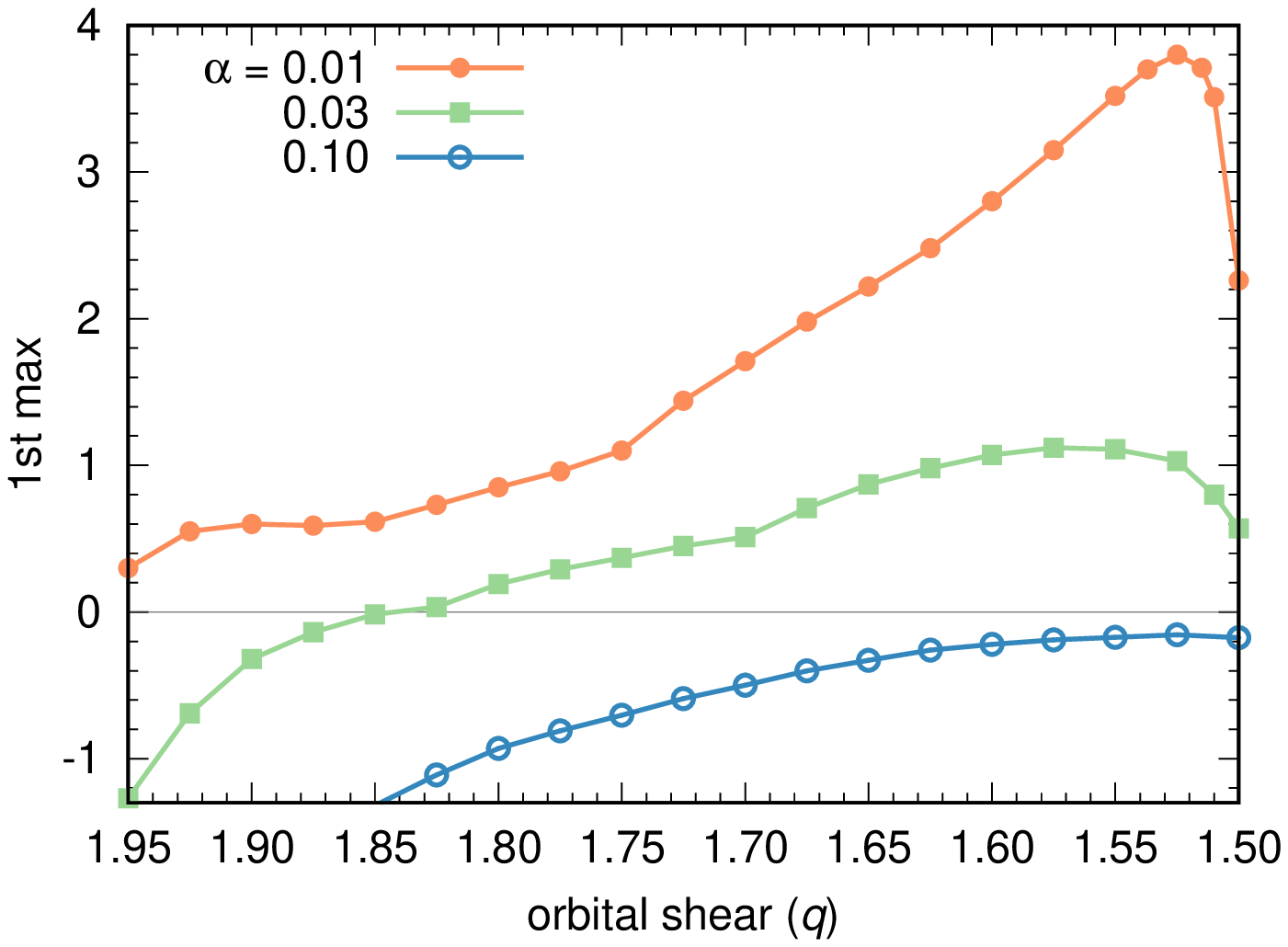}}
    {\includegraphics[angle=270,scale=.55]{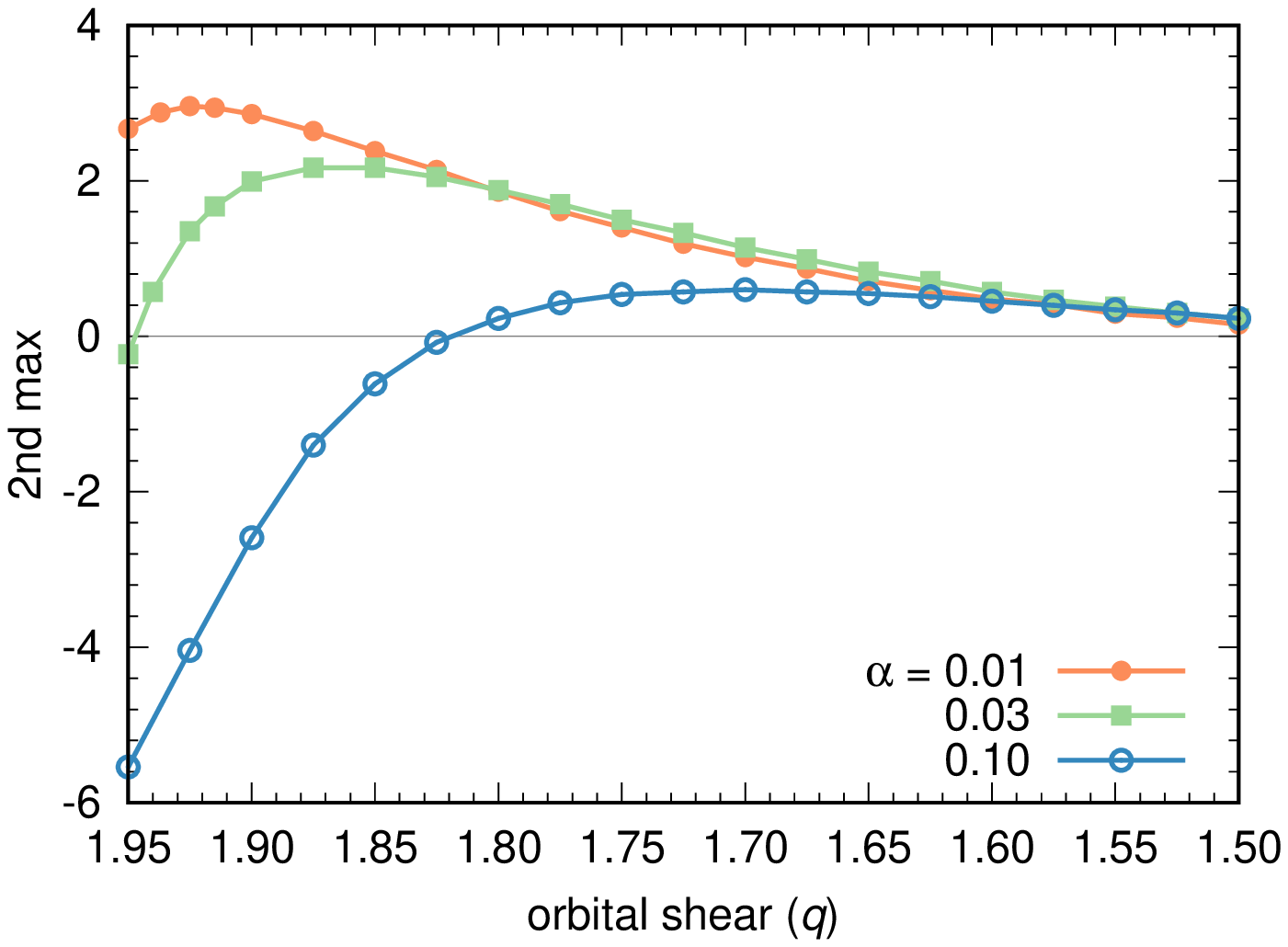}}
    \caption{We present the maximum values of the dimensionless growth rates of instability ($\mathfrak{R}[s]_{\rm max}$) as a function of $q$ for different viscosities. \emph{1st max} corresponds to the maximum growth rate in the low warp amplitude region, $|\psi| \in [0.3,1.0]$, where $Q_2$ term is dominant (left-hand panel), and \emph{2nd max} corresponds to the growth rate which is found at $|\psi| = 5$, i.e. in the region where $Q_1$ term is dominant (right-hand panel). Note that we choose not to plot the 1st max value for $q > 1.85$ in the case of $\alpha=0.1$ as our defined range of $|\psi| \in [0.3,1.0]$ captures the initial down turn from $\Re[s] \approx 0$ at $|\psi|\approx 0.3$ before the behaviour we are trying to exhibit (whereas if we increased the lower bound of the $|\psi|$ range we would miss the peak values for the near-Keplerian cases when $\alpha = 0.01$; cf. Fig. \ref{fig:growthrates}).}
    \label{fig:s_max}
  \end{center}
\end{figure}

In Fig. \ref{fig:s_max}, we present the maximum values of the dimensionless growth rates of instability ($\mathfrak{R}[s]_{\rm max}$) against $q$ for different viscosities. Here, we define two maxima for the growth rates. We see in Fig. \ref{fig:growthrates} that each $\mathfrak{R}[s]$ reaches its peak value somewhere between  $0.3 < |\psi| < 1.0$, where the $Q_2$ term is dominant. We call this peak value \emph{1st max}; specifically we define the \emph{1st max} as the maximum growth rate in the warp amplitude interval $|\psi| \in [0.3,1.0]$. Then, the growth rates show a monotonically increasing behaviour for higher warp amplitudes. We call the growth rate value found at $|\psi| = 5$ as \emph{2nd max}. We see that the values of the \emph{1st max} are highest in the near-Keplerian region. However, we should note that for $\alpha=0.1$ the \emph{1st max} are always negative, implying that the $Q_2$ term is not responsible for instability in such discs whatever the rotation profile they have. On the other hand, the values for the \emph{2nd max} are higher when the disc rotation deviates from Keplerian, except for $\alpha = 0.1$ where strong non-Keplerian rotation acts to stabilize the disc.

\begin{figure}
  \begin{center}
        {\includegraphics[angle=270,scale=.45]{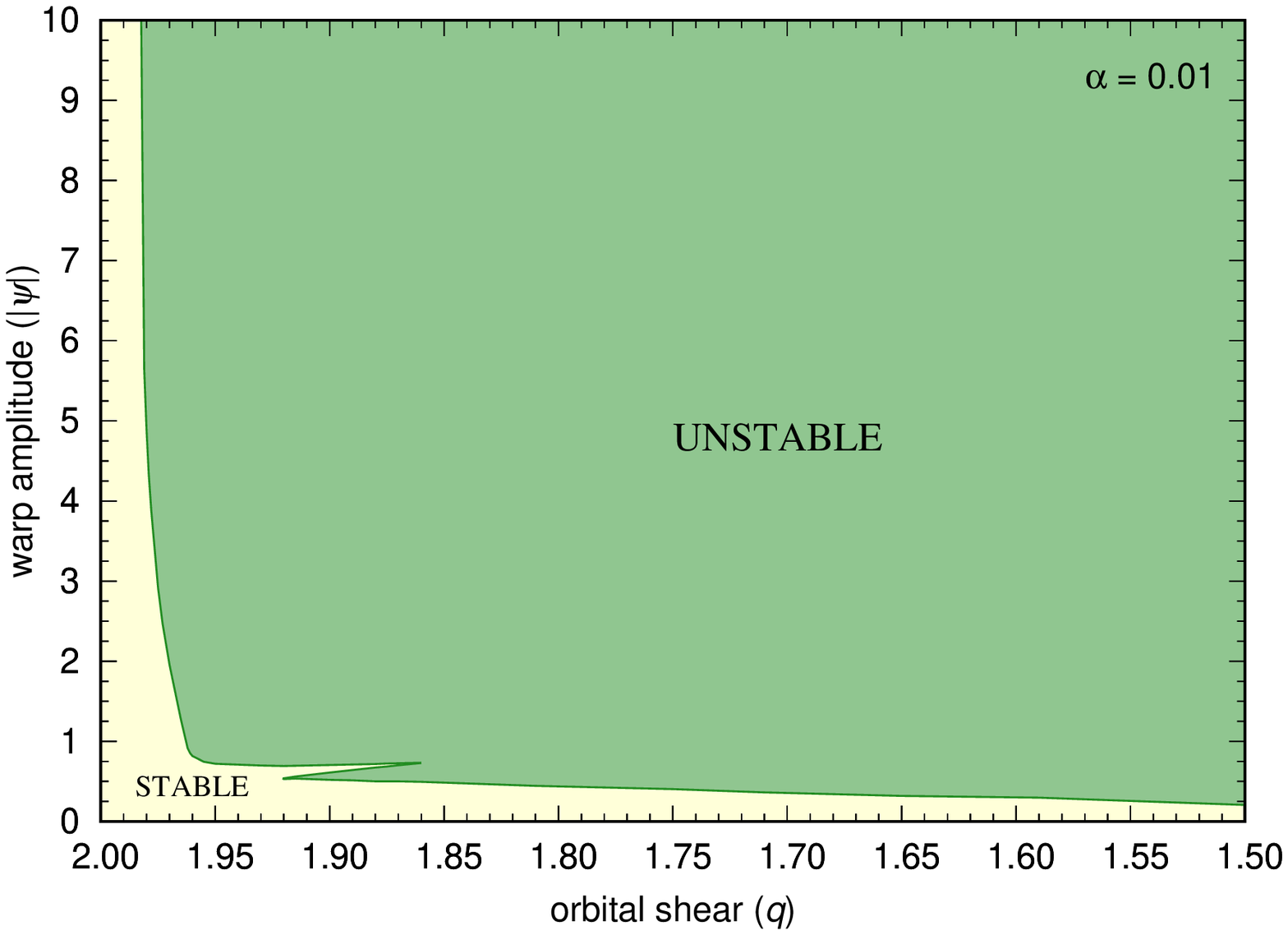}}
        {\includegraphics[angle=270,scale=.45]{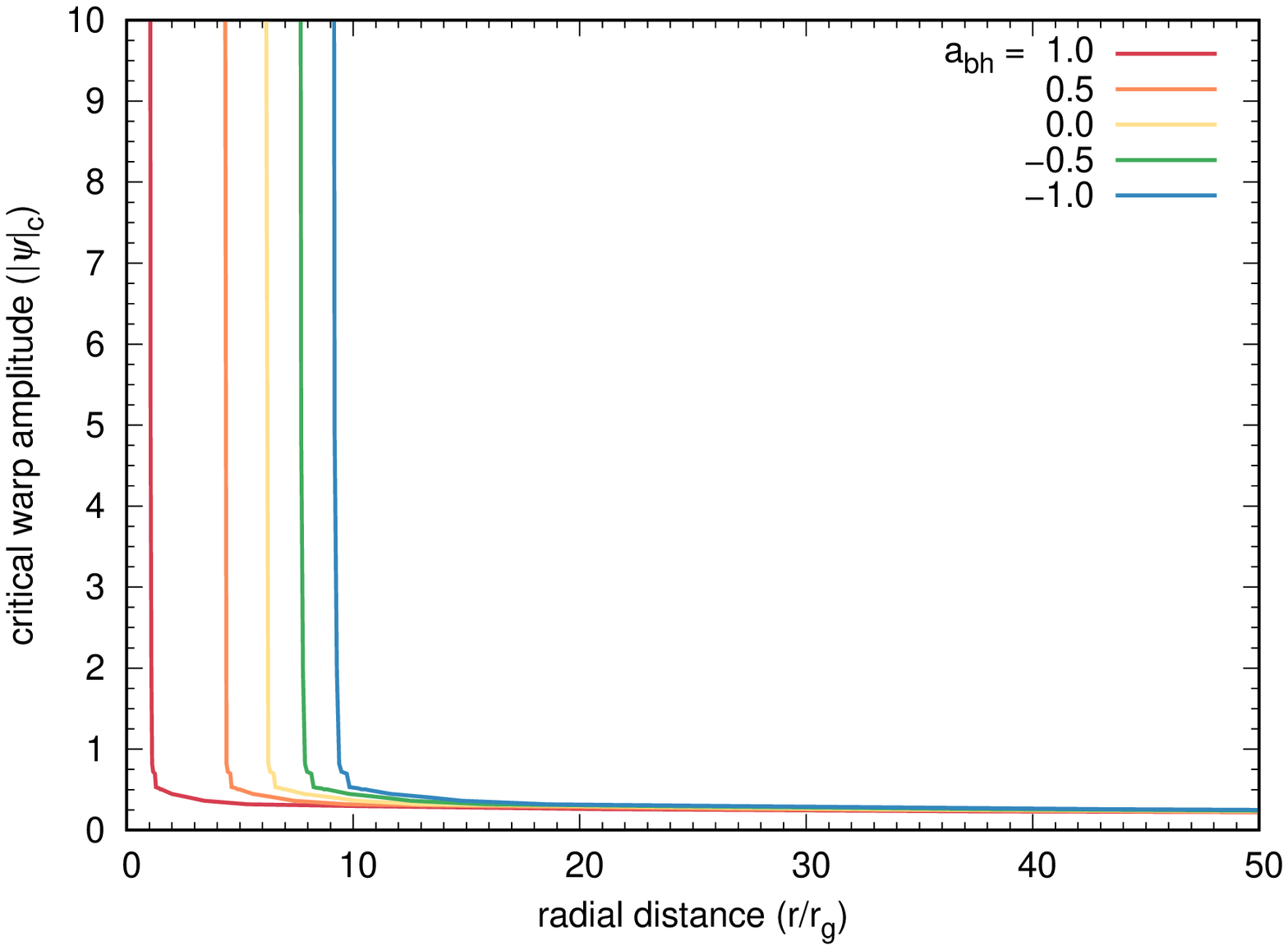}}
        {\includegraphics[angle=270,scale=.45]{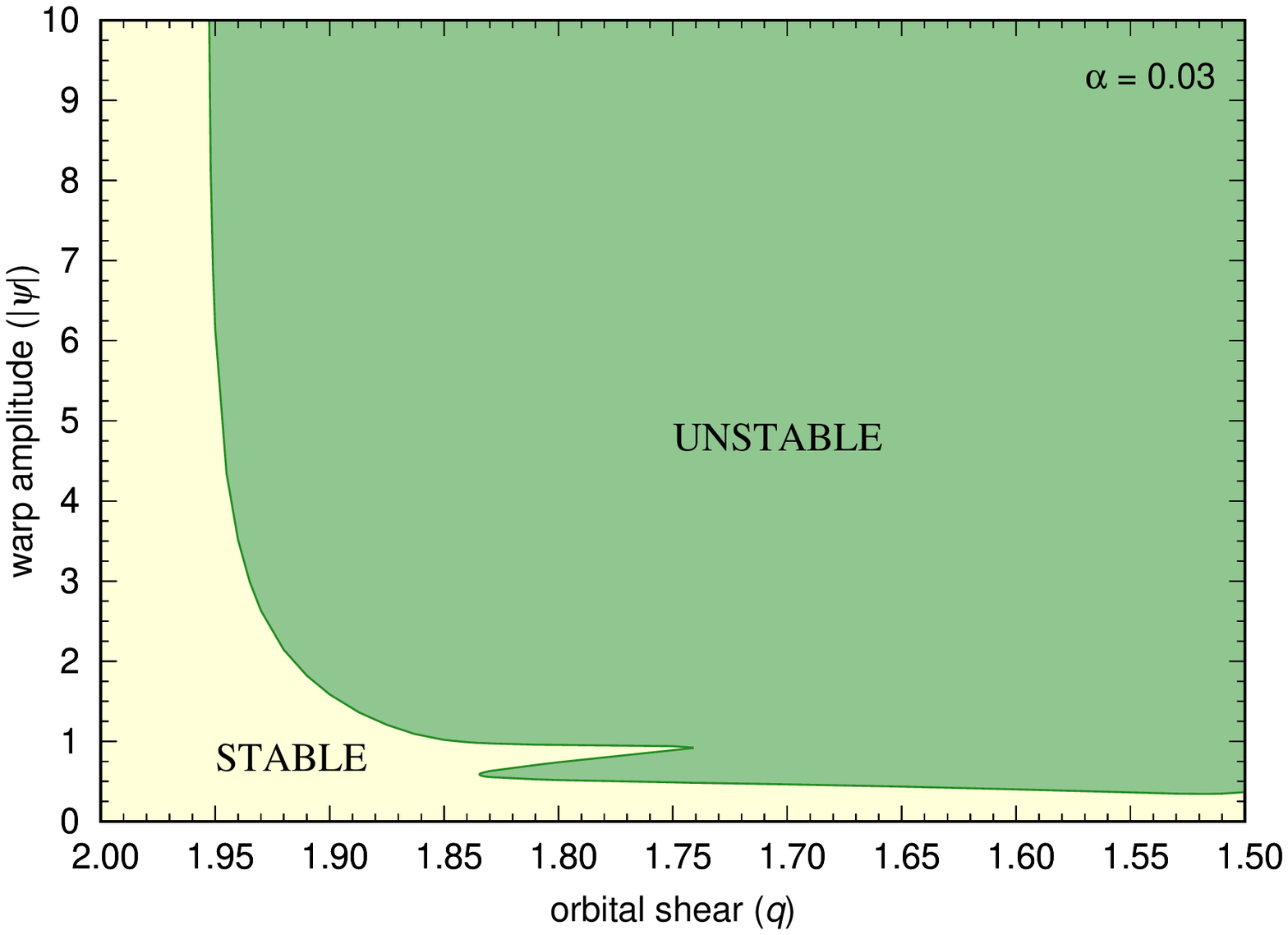}}
        {\includegraphics[angle=270,scale=.45]{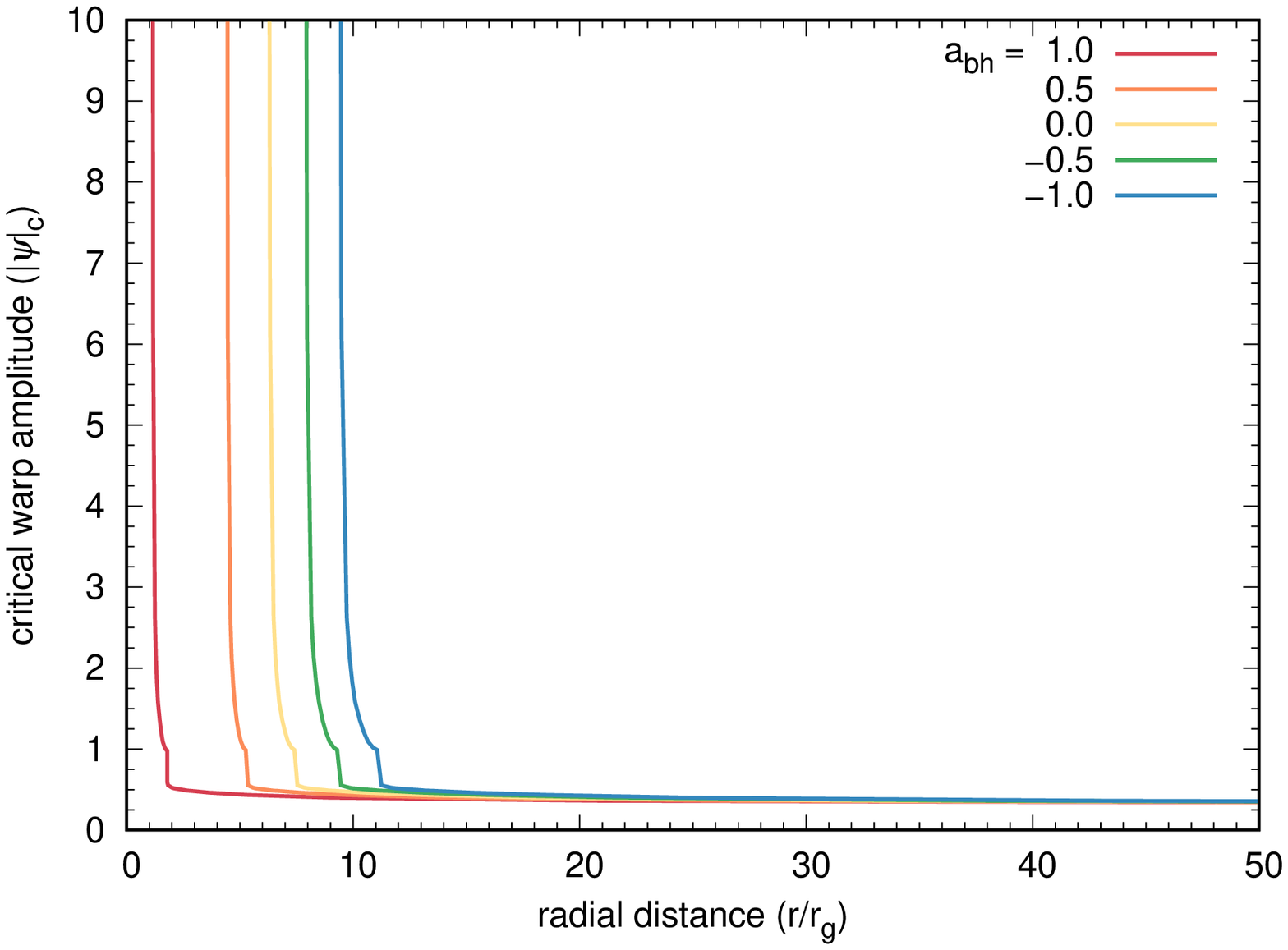}}
        {\includegraphics[angle=270,scale=.45]{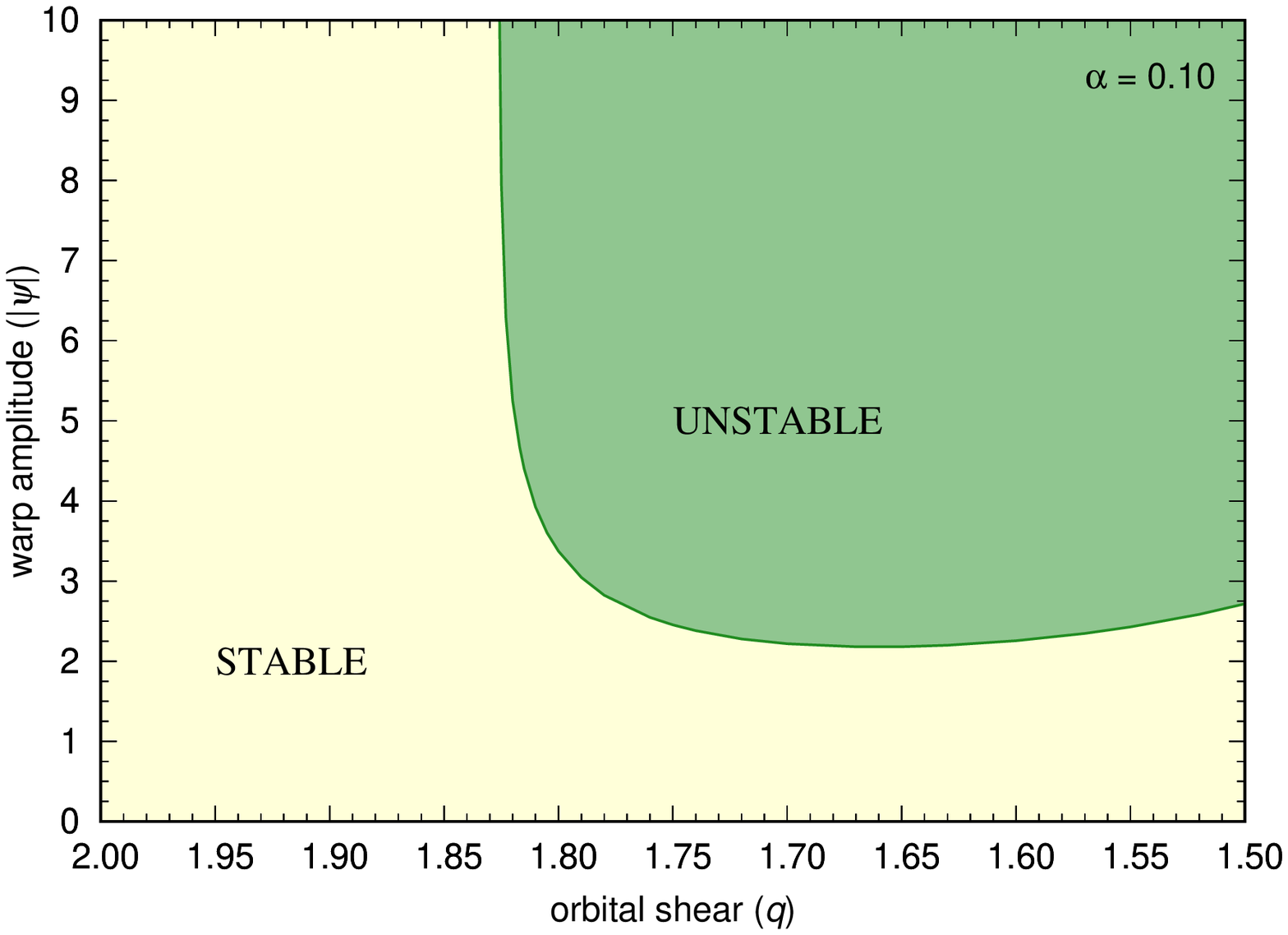}}
        {\includegraphics[angle=270,scale=.45]{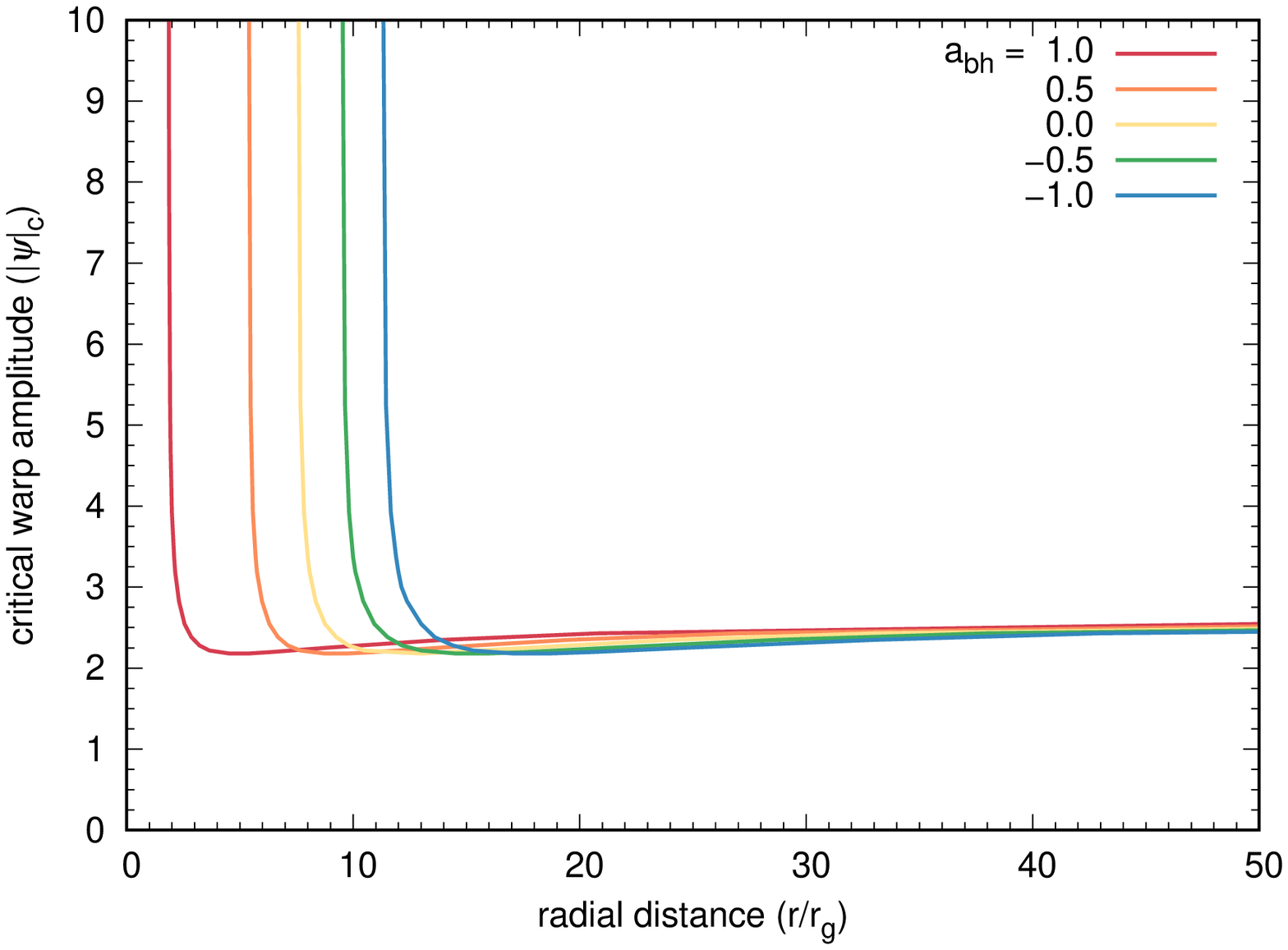}}
        \caption{Left-hand panel: Stable (yellow) and unstable (green) regions in the ($q$,$|\psi|$) parameter space for three different values of the viscosity parameter $\alpha=0.01$ (top), $0.03$ (middle) and $0.1$ (bottom). Each plot shows the critical warp amplitudes for instability to occur in discs with various $q$ values. Right-hand panel: Critical warp amplitudes as a function of ($r/r_{\rm g}$) for various values of the spin parameter ($a_{\rm bh}$) for $\alpha=0.01$ (top), $0.03$ (middle) and $0.1$ (bottom). }
        \label{fig:unstable}
  \end{center}
\end{figure}

\begin{figure}
  \begin{center}
        {\includegraphics[angle=270,scale=.55]{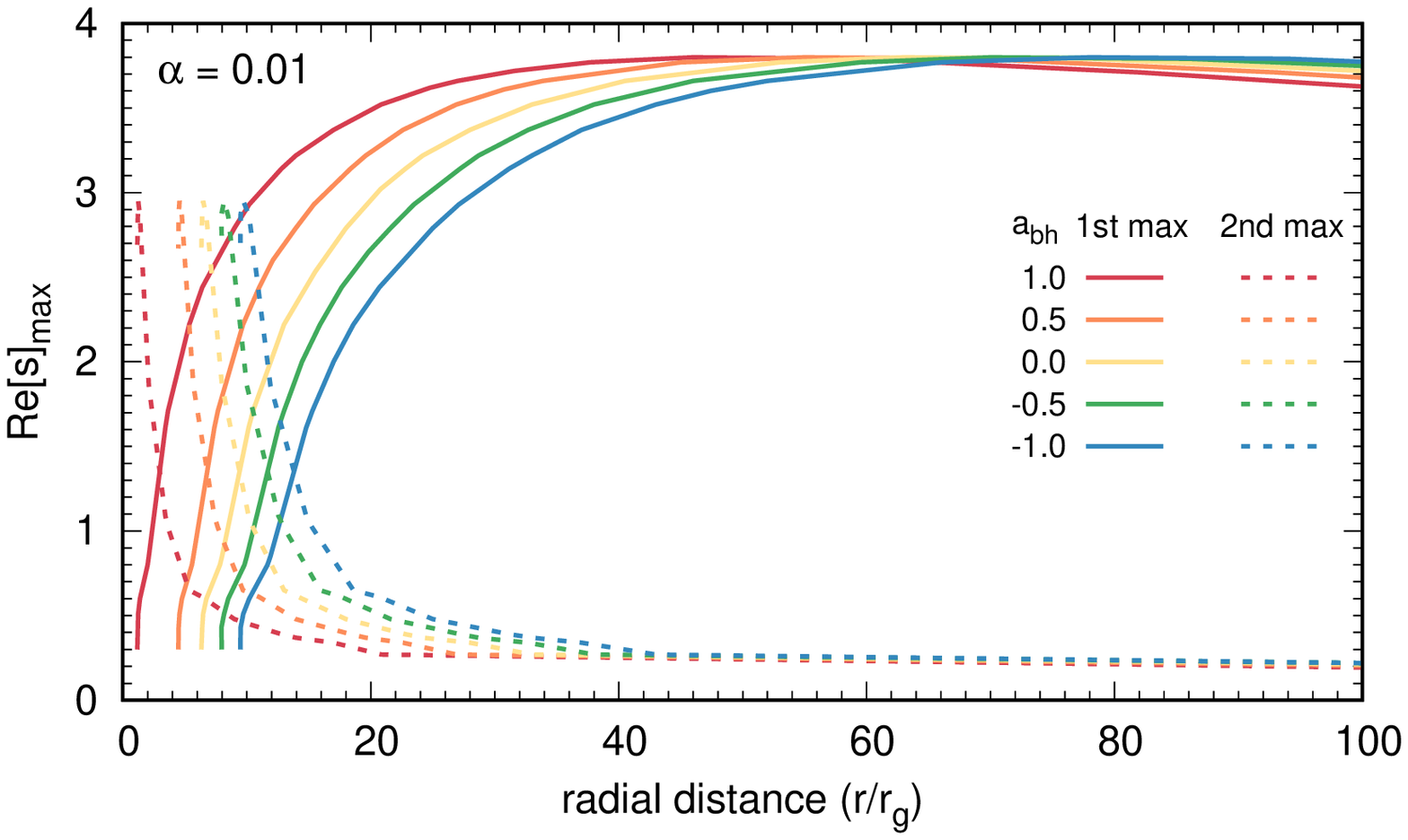}}
         {\includegraphics[angle=270,scale=.55]{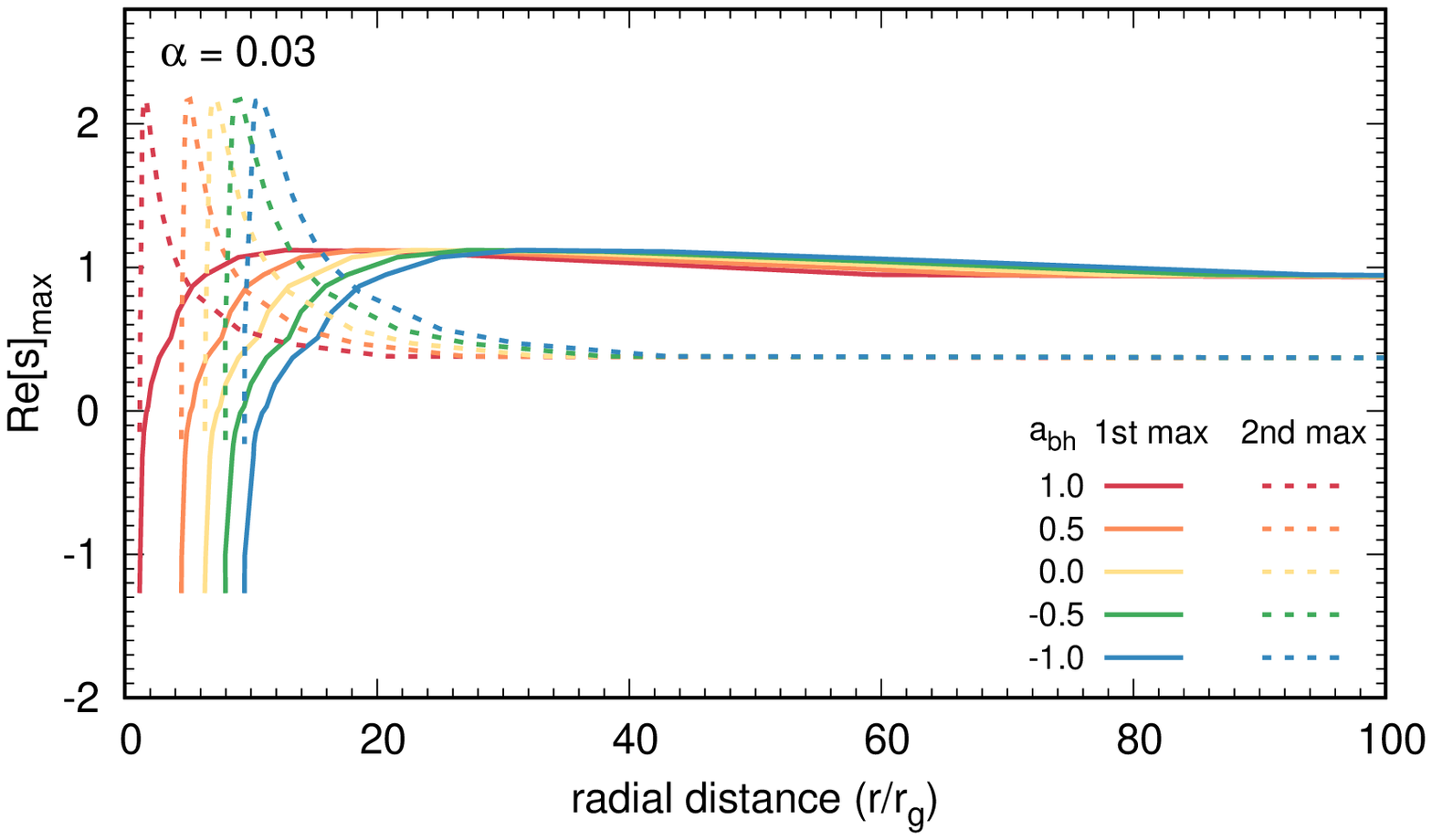}}
          {\includegraphics[angle=270,scale=.55]{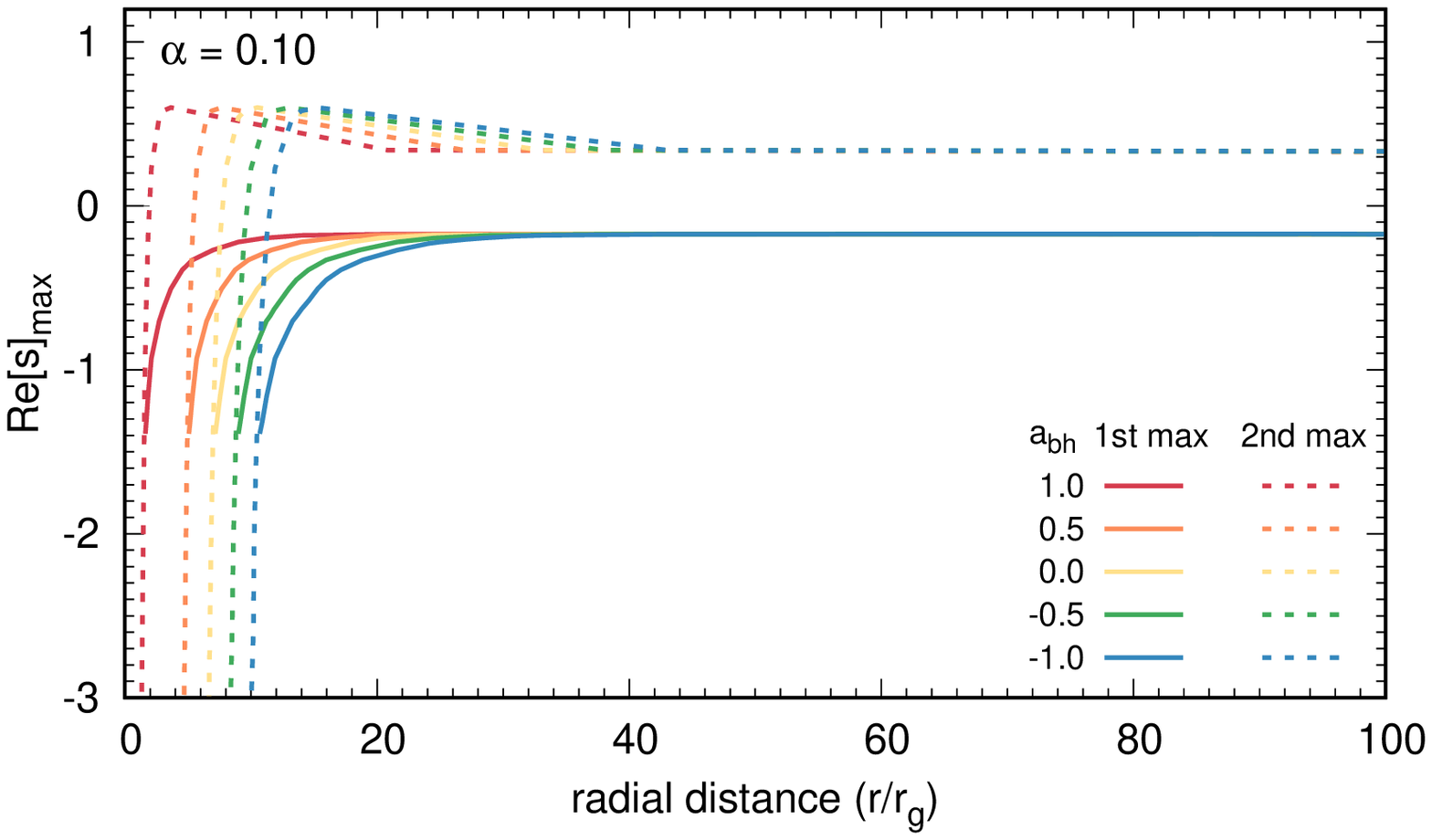}}
          \caption{We present the maximum values of the dimensionless growth rates as a function of radial distance ($r/r_{\rm g}$) for different $\alpha$ values. The continuous line represents \emph{1st max} and the dashed line represents \emph{2nd max}.}
        \label{fig:smax_rg}
  \end{center}
\end{figure}
In Fig. \ref{fig:growthrates}, the warp amplitude value where the dimensionless growth rate becomes positive gives the critical warp amplitude ($|\psi|_c$) for the instability. In the left-hand panel of Fig. \ref{fig:unstable} we show the stable and unstable regions of the $q$--$\left|\psi\right|$ parameter space for different values of $\alpha$. For $\alpha=0.01$ (top panel), we see that instability occurs for all warp amplitudes $|\psi|$ > 0.8 for the dimensionless orbital shear between $ 1.5 < q < 1.96$. For $ 1.96 < q < 1.98$, it is still possible to find a critical warp amplitude which makes the disc unstable. Only the discs with an extremely non-Keplerian rotation profile ($ q \gtrsim 1.98$) are expected to be stable for this case. A nearly flat disc with $|\psi|$ < 0.2 is always stable for $\alpha=0.01$. Similarly, for $\alpha=0.03$ (middle panel), the disc becomes unstable for all warp amplitudes $|\psi|$ > 1 when the orbital shear lies between $ 1.5 < q < 1.85$. For $1.8 < q < 1.95 $, a higher warp amplitude is needed for instability. The discs with $q \gtrsim 1.95$ are always stable for this case. The discs with $|\psi| \lesssim 0.4$ are also always stable. Finally, for $\alpha=0.1$ (bottom panel), the critical warp amplitudes which give rise to instability are found to be higher than those we find for low viscosities. The instability requires warp amplitudes of $|\psi| \gtrsim 2.5 - 3$ for $ 1.5 < q < 1.8$. The discs with $q\gtrsim 1.825$ are expected to be stable for all warp amplitudes. Discs with $|\psi|$ < 2.2 are always stable for $\alpha=0.1$.

We find that for a given value of $\alpha$, the critical warp amplitudes are smaller for Keplerian and near-Keplerian orbits. However the growth rates are complicated in that at small warp amplitudes (where the $Q_2$ term dominates) the growth rates are higher for near-Keplerian rotation, while at large warp amplitudes (where the $Q_1$ term dominates) they are higher for non-Keplerian rotation (e.g. left and middle panel of Fig.~\ref{fig:growthrates}); however, for larger values of $\alpha$ (e.g. right hand panel of Fig.~\ref{fig:growthrates}) large $q$ can stabilize the disc for all values of $\left|\psi\right|$. This is shown in Fig.~\ref{fig:s_max} where the left hand panel (for small $\left|\psi\right|$) shows lower growth rates as $q$ increases, while the right hand panel (for larger $\left|\psi\right|$) shows increasing growth rates for increasing $q$ except where $q\gtrsim 1.9$ or where $\alpha = 0.1$. Thus, while in detail the evolution is quite parameter dependent, we find that in general higher viscosity discs are more stable (higher critical warp amplitudes and lower growth rates). We also find that non-Keplerian rotation provides a stabilizing effect at low warp amplitudes, but at high warp amplitudes non-Keplerian rotation can increase the growth rate if the disc is unstable. We plot the \emph{1st max} (defined as the maximum growth rate in the warp amplitude interval $|\psi| \in [0.3,1.0]$) and \emph{2nd max} (defined as the growth rate at a warp amplitude of $|\psi| = 5$) growth rates as a function of radial distance from the black hole for different viscosities and spin parameters in Fig.~\ref{fig:smax_rg}.

\subsection{Validity of $\kappa-q$ relation for black hole discs}
\begin{figure}
  \begin{center}
        {\includegraphics[angle=270,scale=.50]{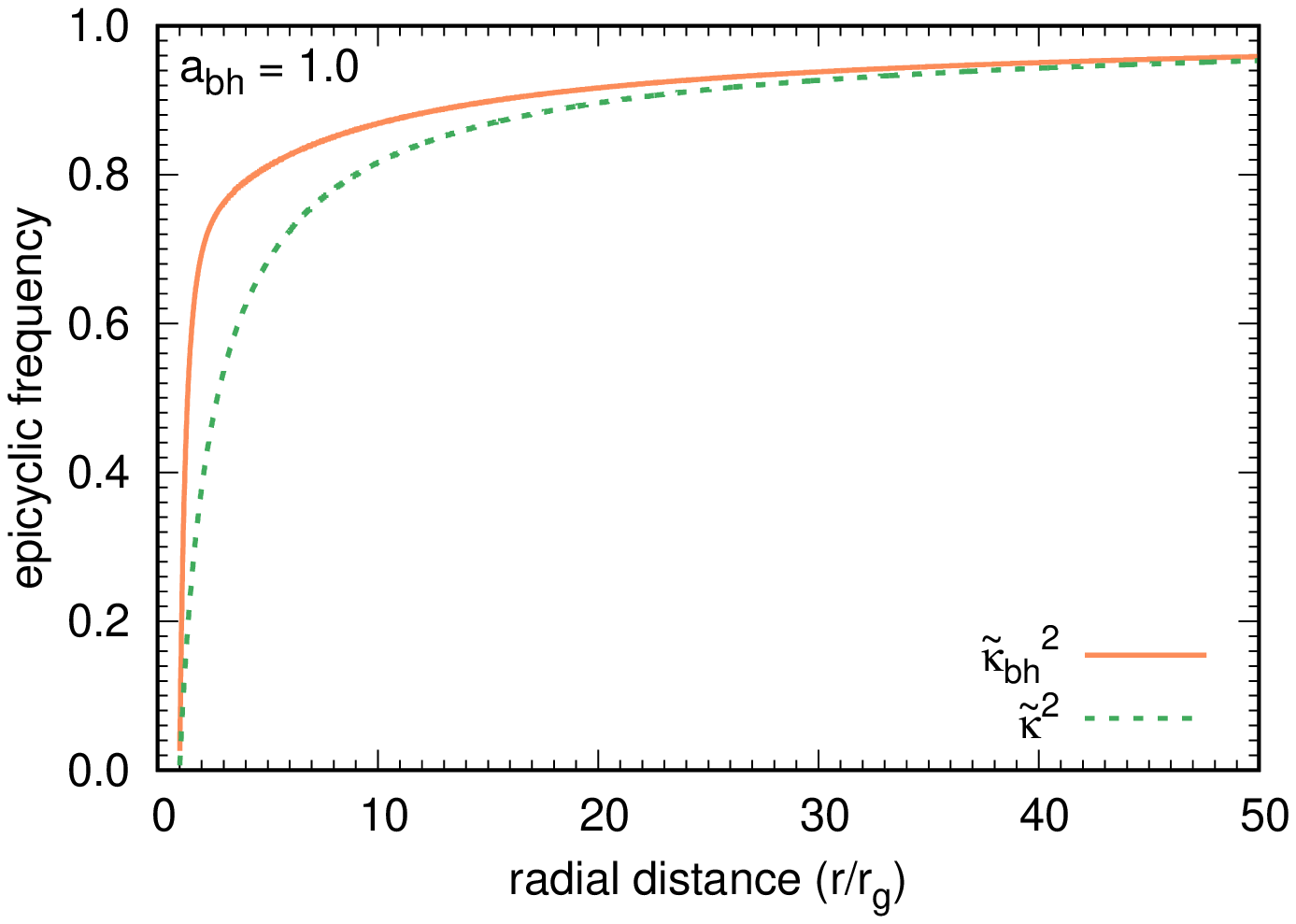}}
         {\includegraphics[angle=270,scale=.50]{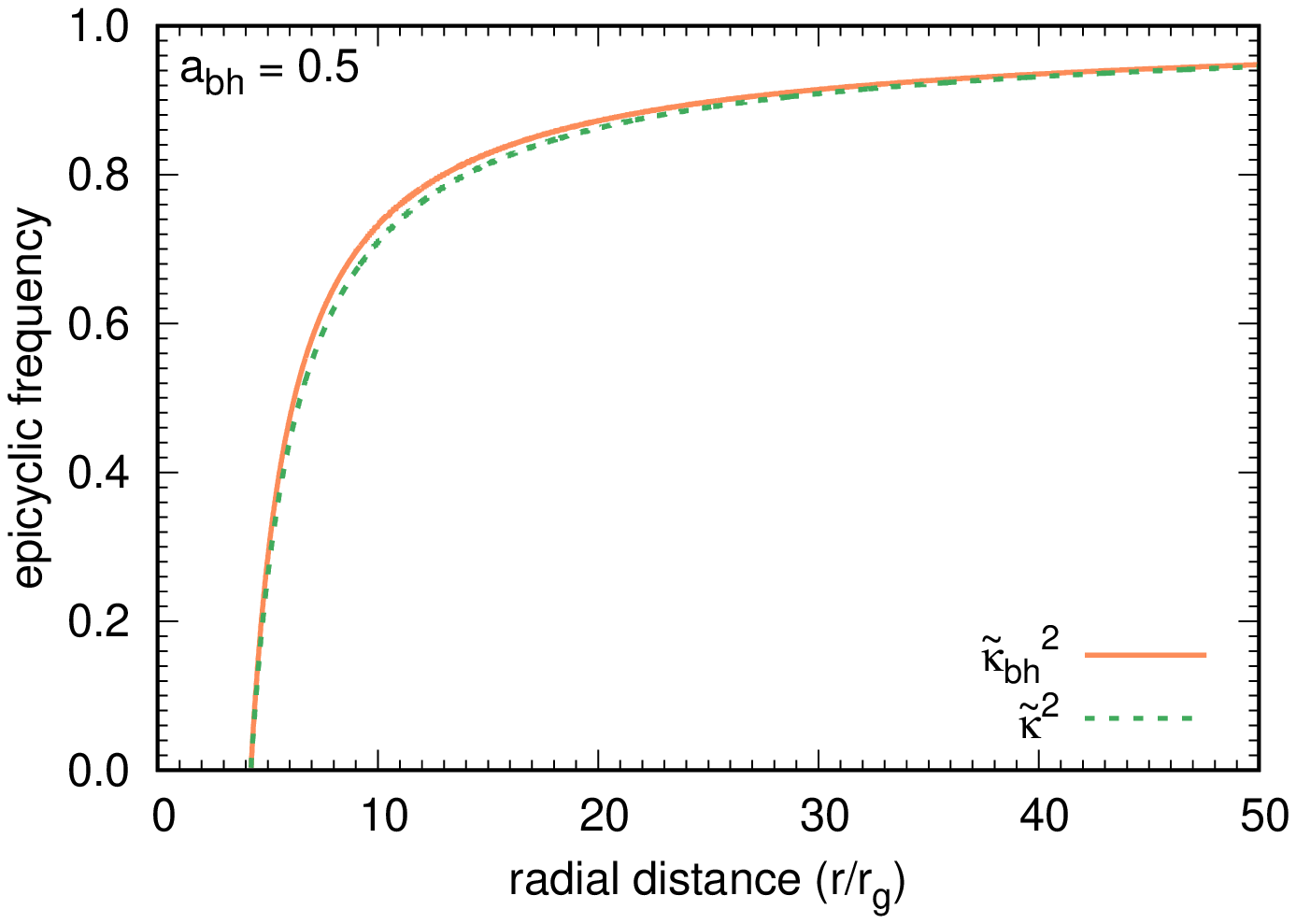}}
          {\includegraphics[angle=270,scale=.50]{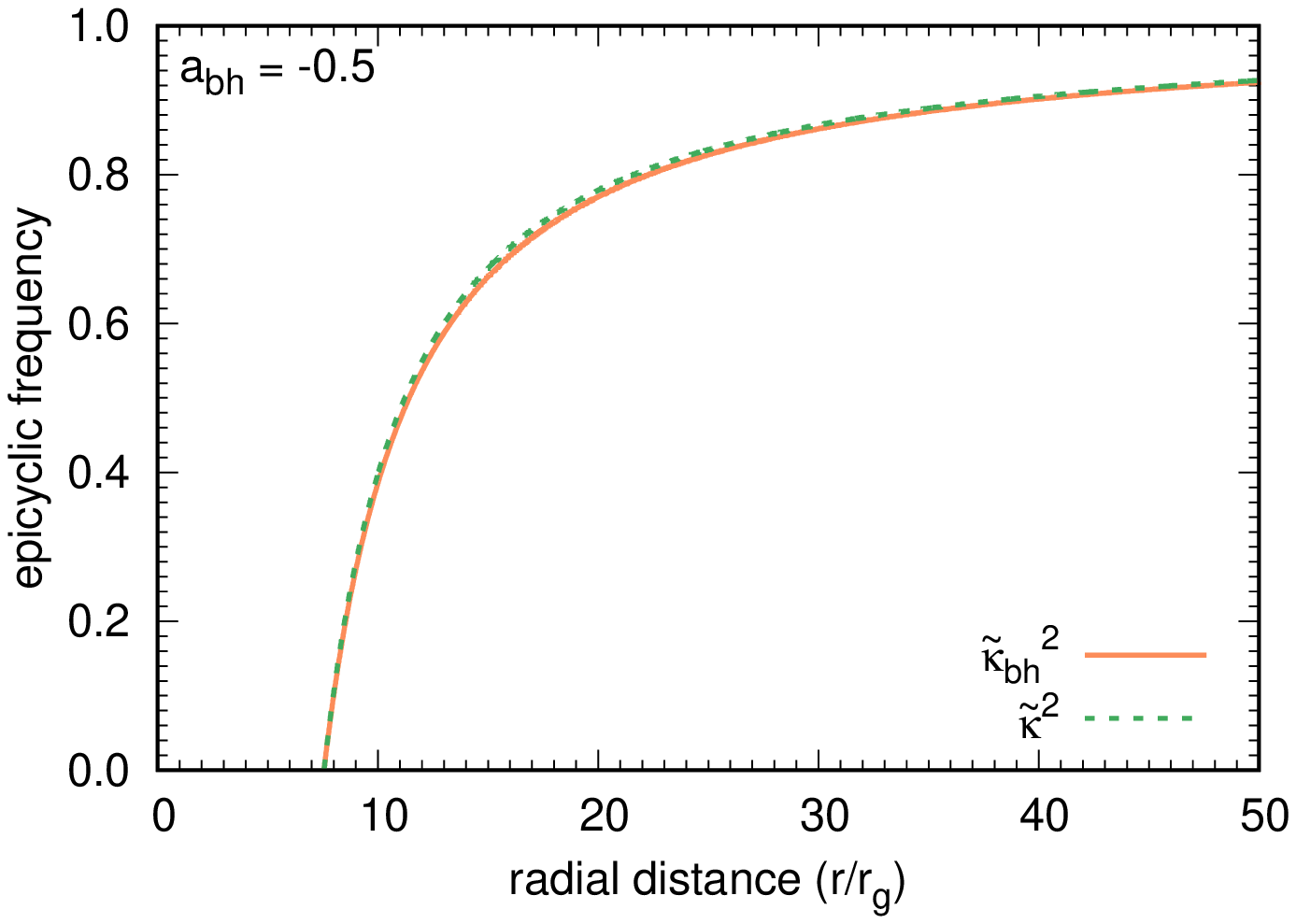}}
              {\includegraphics[angle=270,scale=.50]{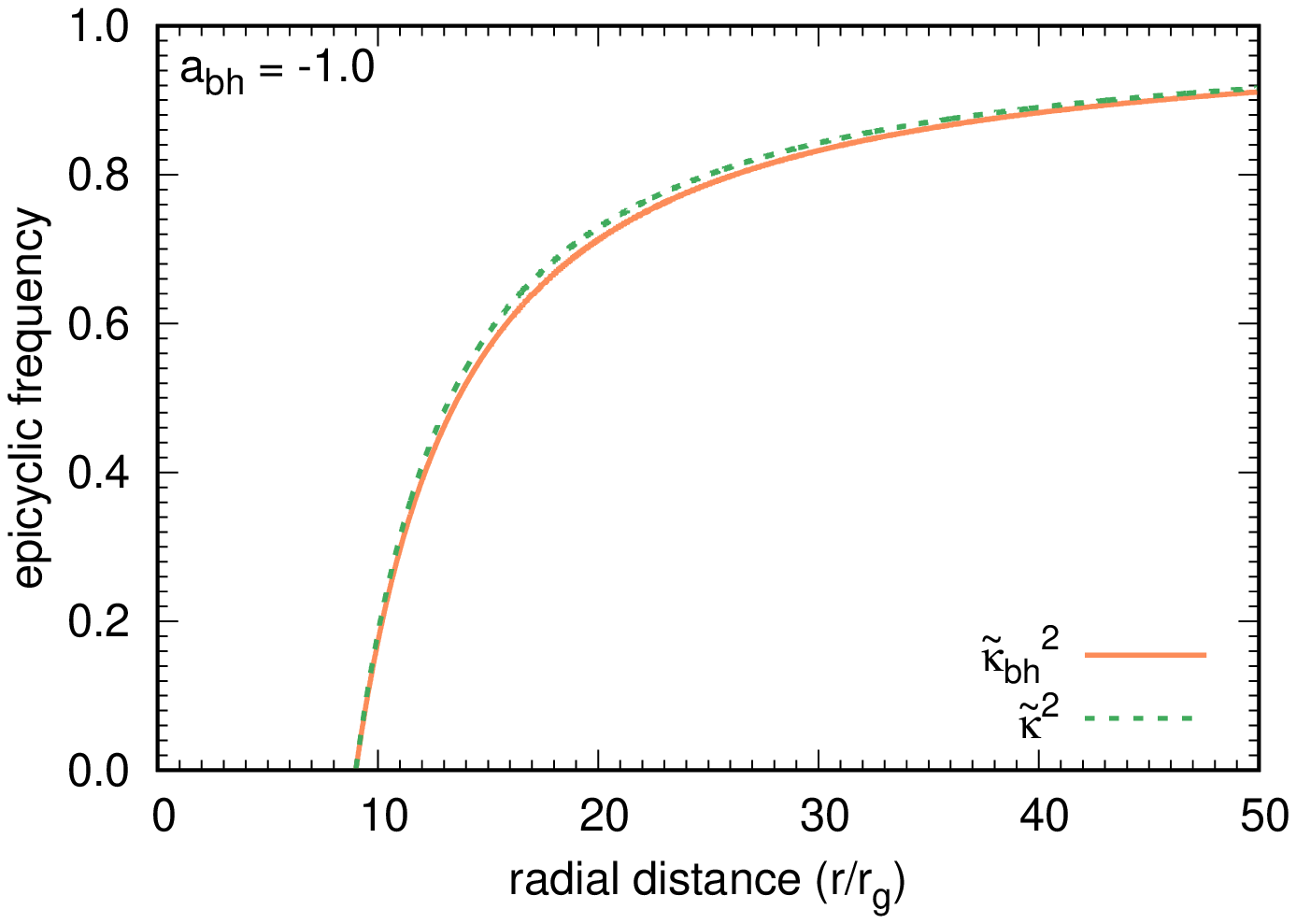}}
              \caption{We compare the dimensionless epicyclic frequency values calculated by writing $q_{\rm bh}$ in (\ref{eq:kappa_q_bh}) which correspond to $\tilde{\kappa}^2_{\rm bh}$ given by (\ref{eq:kappa_bh}), and calculated by writing $q_{\rm bh}$ in (\ref{eq:kappa_q}) which corresponds to the assumption made in evaluating the torque coefficients.}
        \label{fig:appendix1}
  \end{center}
\end{figure}

In both the stability analysis and the calculation of the torque coefficients we make use of the relation between epicyclic and orbital frequency given by equation~\ref{eq:kappa_q}. However, as discussed in Section~\ref{sec:nonkep1} this relation does not hold exactly for the frequencies quoted by \cite{Gammie:2004,Penna:2013} for discs around black holes. Far from the black hole, or for Schwarzschild black holes the relation holds. Thus the values quoted above should not be taken as precise for radii close to the ISCO. We can provide an indication of the level of error introduced by this by comparing the values of $\tilde{\kappa}^2$ that are calculated by (i) putting $q_{\rm bh}$ into (\ref{eq:kappa_q_bh}) which correspond to (\ref{eq:kappa_bh}), and (ii) calculated by putting $q_{\rm bh}$ in (\ref{eq:kappa_q}). We plot these in Fig.~\ref{fig:appendix1}, and we can see that the dimensionless epicyclic frequency values found from (\ref{eq:kappa_q_bh}) slightly differ from those found from (\ref{eq:kappa_q}) near the black hole. The deviation is remarkable only when $a_{\rm bh} \sim 1$. We expect the use of equation~\ref{eq:kappa_q} would slightly shift the corresponding distances to the values of the critical warp amplitudes presented in the right panel of Fig. \ref{fig:unstable} and to the maximum growth rates presented in Fig.~\ref{fig:smax_rg}. However, the change is expected to be small, and does not affect our main results.

\section{Conclusion}
We have provided an analysis of the effects of non-Keplerian rotation on the stability of warped discs where $\alpha > H/R$, that is, the case where the warp is propagated principally by local diffusion rather than wave transport. We find that, as for the case of Keplerian rotation, higher viscosity discs are generally more stable with a higher critical warp amplitude and lower growth rates than lower viscosity discs. We also find that as the rotation deviates more strongly from Keplerian, the disc is increasingly stabilized at small warp amplitudes, while at large warp amplitudes the growth rates can be higher. The strong dependence of the evolution of these discs on the value of $\alpha$, the shear rate $q$, and the warp amplitude $\left|\psi\right|$, is likely to be responsible for the rich dynamics found in numerical simulations of these objects.

For discs subject to tides from a binary system, e.g. circumstellar discs with a companion or circumbinary discs, we find that the departures from Keplerian rotation are small enough that the general picture from a Keplerian analysis is still valid, as the growth rates and critical warp amplitudes are similar to the Keplerian case. These types of disc may occur in stellar binary systems, such as CVs or X-ray binaries. In these cases the discs are expected to be thin enough, and the viscosity high enough, that warps propagate diffusively. The outer regions of these discs (or the inner regions of circumbinary discs) are where tides are strongest, but even here the departure from Keplerian rotation is still quite small as the discs are efficiently truncated by orbital resonances \citep{Papaloizou:1977aa,Artymowicz:1994aa}. Protoplanetary discs are also expected to have rotation that is close to Keplerian, but our analysis cannot be directly applied to protoplanetary discs as these discs are likely to propagate warps with a significant wavelike component, as $H/R$ is typical greater than $\alpha$. 

However, we find that non-Keplerian rotation has a significant impact on the dynamics of a warp in the inner regions of discs around black holes. The sharp upturn in critical warp amplitude at small radii seen in Fig.~\ref{fig:unstable} suggests that discs may not break all the way down to the ISCO, instead the disc is expected to rapidly align to the black hole spin in these regions. However, for lower values of the viscosity $\lesssim 0.1$, the critical warp amplitudes rise significantly only close to the ISCO, and thus it should be possible for the disc to break into discrete rings that can precess independently close to the black hole. We note that this is a requirement of the Relativistic Precession Model (RPM; \citealt{Stella:1998aa,Ingram:2010aa,Motta:2014aa,Ingram:2014aa,Motta:2018aa}) which assumes that some of the frequencies observed in the power spectra of light curves from accreting compact objects are caused by nodal and apsidal precession of disc material at discrete radii. It is, of course, not possible for a fluid disc to exhibit repeated {\it local} precession at the driving frequency (e.g. Lense-Thirring precession frequency) coherently over many cycles without the disc breaking into discrete parts (if the disc cannot break, a non-local response such as a propagating warp occurs). Therefore we conclude that if the observed frequencies are indeed caused by relativistic nodal and apsidal precession, then it is consistent with and requires that the disc be unstable to breaking into discrete rings (i.e. the disc tearing behaviour suggested by \citealt{Nixon:2012ad}). It may then be possible to understand the wide range of complex variability observed in accreting black hole (and neutron star) sources with a model that takes into account the dependence of the disc warping instability discussed here on the local parameters of the accretion disc throughout the accretion cycle \citep[cf.][]{Nixon:2014aa}.

\section*{Acknowledgments}
We thank the referee for useful comments. We thank Gordon Ogilvie for providing the code to calculate the torque coefficients. We thank Jim Pringle and Eric Coughlin for useful discussions. SD acknowledges the warm hospitality of the Department of Physics and Astronomy at University of Leicester during her visit. SD is supported by the Turkish Scientific and Technical Research Council (T\"{U}B\.{I}TAK - 117F280). CJN is supported by the Science and Technology Facilities Council (grant number ST/M005917/1).

\bibliographystyle{mnras}
\bibliography{nixon}

\bsp
\label{lastpage}
\end{document}